\documentclass[pra,aps,showpacs,groupedaddress,superscriptaddress,twocolumn,toc=flat,longbibliography]{revtex4-1}

\usepackage[colorlinks=true,linkcolor=blue,urlcolor=blue,citecolor=blue]{hyperref}

\usepackage[utf8x]{inputenc}
\usepackage{color}
\usepackage{bbm} 

\usepackage{amsfonts,ams math,amssymb,stmaryrd}

\usepackage{graphicx}
\usepackage{subfigure}  
\usepackage{bbm} 
\usepackage{hyperref}
\usepackage{epsfig}
\usepackage{mathrsfs}
\usepackage{verbatim}
\usepackage{centernot}
\usepackage{ulem}
\usepackage{mathtools}
\usepackage{nicefrac}

\renewcommand{\l}{\left(}
\renewcommand{\r}{\right)}

\newcommand{\bra}[1]{\langle#1|}
\newcommand{\ket}[1]{|#1\rangle}

\renewcommand{\ij}{{\langle i, j \rangle}}
\renewcommand{\H}{\hat{\mathcal{H}}}

\renewcommand{\c}{\hat{c}}
\newcommand{\f}{\hat{f}}
\newcommand{\tf}{\hat{\tilde{f}}}
\newcommand{\tfd}{\hat{\tilde{f}}^\dagger}

\newcommand{\cd}{\hat{c}^\dagger}

\newcommand{\bd}{\hat{b}^\dagger}

\renewcommand{\b}{\hat{b}}
\newcommand{\hd}{\hat{h}^\dagger}
\newcommand{\h}{\hat{h}}

\newcommand{\n}{\hat{n}}

\newcommand{\hc}{\text{h.c.}}

\newcommand{\psd}{\hat{\psi}^\dagger}
\newcommand{\ps}{\hat{\psi}}

\newcommand{\fd}{\hat{f}^\dagger}

\usepackage{array}

\usepackage{cancel,ifthen}
\newcommand{\cmnt}[2][NoInPuT]{\ifthenelse{\equal{#1}{NoInPuT}}{}{{\color{red}\sout{#1}}} {\color{blue} #2}}

\usepackage{bm}	
\renewcommand{\vec}[1]{\bm{#1}}

\begin{document}
\normalem	

\title{Angle-resolved photoemission spectroscopy with quantum gas microscopes}

\author{A. Bohrdt}
\affiliation{Department of Physics and Institute for Advanced Study, Technical University of Munich, 85748 Garching, Germany}
\affiliation{Department of Physics, Harvard University, Cambridge, Massachusetts 02138, USA}

\author{D. Greif}
\affiliation{Department of Physics, Harvard University, Cambridge, Massachusetts 02138, USA}

\author{E. Demler}
\affiliation{Department of Physics, Harvard University, Cambridge, Massachusetts 02138, USA}

\author{M. Knap}
\affiliation{Department of Physics and Institute for Advanced Study, Technical University of Munich, 85748 Garching, Germany}

\author{F. Grusdt}
\affiliation{Department of Physics, Harvard University, Cambridge, Massachusetts 02138, USA}


\date{\today}

\begin{abstract}

Quantum gas microscopes are a promising tool to study interacting quantum many-body systems and bridge the gap between theoretical models and real materials. So far they were limited to measurements of instantaneous correlation functions of the form $\langle \hat{O}(t) \rangle$, even though extensions to frequency-resolved response functions $\langle \hat{O}(t) \hat{O}(0) \rangle$ would provide important information about the elementary excitations in a many-body system. For example, single particle spectral functions, which are usually measured using photoemission experiments in electron systems, contain direct information about fractionalization and the quasiparticle excitation spectrum. Here, we propose a measurement scheme to experimentally access the momentum and energy resolved spectral function in a quantum gas microscope with currently available techniques. As an example for possible applications, we numerically calculate the spectrum of a single hole excitation in one-dimensional $t-J$ models with isotropic and anisotropic antiferromagnetic couplings. A sharp asymmetry in the distribution of spectral weight 
appears when a hole is created in an isotropic Heisenberg spin chain. This effect slowly vanishes for anisotropic spin interactions and disappears completely in the case of pure Ising interactions. The asymmetry strongly depends on the total magnetization of the spin chain, which can be tuned in experiments with quantum gas microscopes. An intuitive picture for the observed behavior is provided by a slave-fermion mean field theory. The key properties of the spectra are visible at currently accessible temperatures.
\end{abstract}

\maketitle

\section{Introduction}
\label{secIntro}

Ultracold atomic gases provide a versatile platform to study quantum many-body physics from a new perspective. They enable insights into systems that are on one hand challenging to describe theoretically and on the other hand difficult to realize with a comparable amount of isolation, control, and tunability in solid state systems. Recently, we have seen dramatic progress in the quantum simulation of the Fermi-Hubbard model, which in 2D is believed to capture essential features of high-temperature cuprate superconductors \cite{Hofstetter2002,Greif2013,Hart2015}. Experimental results from quantum gas microscopy of ultracold fermions in optical lattices \cite{Parsons2015,Omran2015,Cheuk2015,Haller2015,Parsons2016,Boll2016,Cheuk2016,Mitra2017} have already demonstrated spin-charge separation in one-dimensional (1D) systems \cite{Hilker2017} as well as long-range anti-ferromagnetic correlations \cite{Mazurenko2016} and canted antiferromagnet states \cite{Brown2017} in two dimensions. In order to relate cold atom experiments to their solid state counterparts and facilitate direct comparisons, it is desirable to measure similar physical observables in both systems \cite{Chin2004,Kollath2007,Dao2007,Hart2015,Knap2012,Knap2013,Cetina2016}.

\begin{figure}[t!]
\centering
\epsfig{file=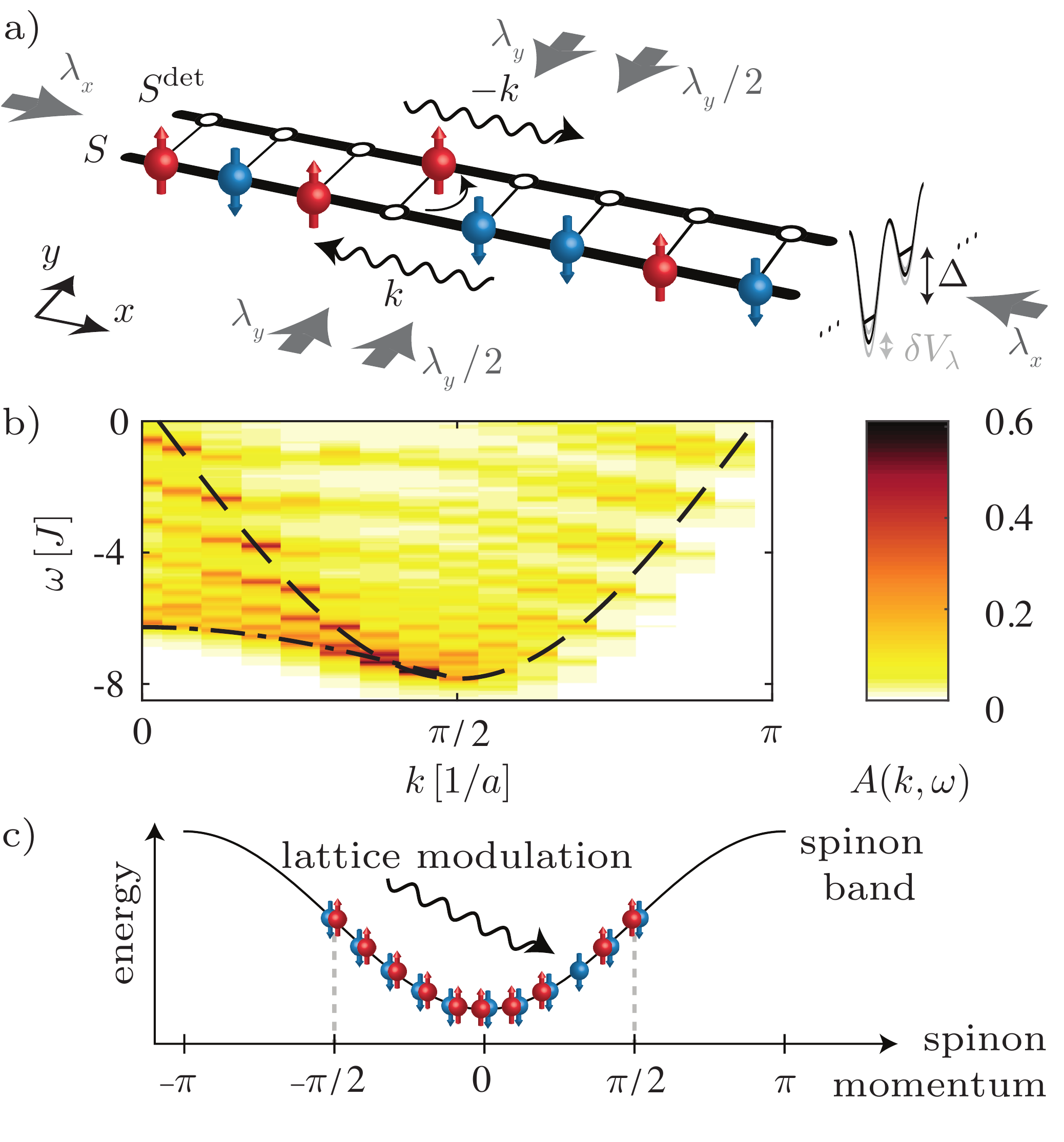, width=\columnwidth}
\caption{\textbf{Measuring the single-hole spectral function.} a)~Proposed experimental setup. A lattice modulation along the $y$-direction creates a hole in the physical system $S$ by transferring a single particle into the neighboring, thermodynamically disconnected detection system $S^{\mathrm{det}}$, which is offset in energy by $\Delta$. A subsequent momentum space mapping technique enables the determination of the momentum $k$ of the excitation. The rate of the transferred atoms is proportional to the spectral function $A(k,\omega)$. b)~Exemplary calculated spectral function of the $t-J$ model with next-nearest neighbor interactions and isotropic spin couplings for $L=16$ sites, tunneling $t/J=4$, temperature $T/J = 0.2$ and open boundary conditions. The spectral weight in units of $1/J$ is color coded. Individual holon and spinon branches in the spectrum are clearly visible, as indicated by the dashed and dashed-dotted lines. c)~In a mean field approach, the ground state of the effective spin degrees of freedom, which is a Luttinger spin liquid, is described as a half-filled Fermi sea of spinons. In the measurement process, a holon is created and a spinon is removed, such that the accessible momenta are restricted to $k \lesssim \pi/2$ at low energies, which explains the asymmetry in b).
}
\label{fig1Setup}
\end{figure}

Traditional solid state experiments rely on measurements of time-dependent response functions of the form $\langle \hat{O}(t) \hat{O}(0) \rangle$ in the frequency domain~\cite{retardedGF}. Examples include inelastic neutron scattering, X-ray spectroscopy, scanning tunneling microscopy, angle resolved photoemission spectroscopy (ARPES), or purely optical probes. In contrast, quantum gas microscopes are used to perform destructive measurements accompanied by a collapse of the many-body wavefunction. While this gives immediate access to instantaneous correlation functions of the form $\langle \hat{O}_1(t)\hat{O}_2(t) ... \hat{O}_n(t) \rangle$, extensions to frequency-resolved response functions have not been realized so far.

One of the most powerful tools for studying strongly correlated electrons in solids is angle resolved photoemission spectroscopy (ARPES). In this technique, electrons are ejected from the surface of a sample through the photoelectric effect. By counting the number of photoelectrons and measuring their energy $\omega$ and momentum $k$, the single-particle excitation spectrum $A(k,\omega)$ is obtained. The spectral function reveals fundamental properties of the system and its excitations \cite{Damascelli2003,Lee2006}, and important insights about high-$T_c$ cuprate superconductors have been obtained from ARPES measurements. One of the most puzzling observations in this context is the appearance of Fermi arcs in the spectrum below optimal doping in the pseudogap phase \cite{Lee2006}. A microscopic understanding of this phenomenon is currently lacking, and it is expected that experiments with ultracold atoms can shed new light on this long-standing problem.

Spectral functions have already been measured in fermionic quantum gas experiments for instance by radio-frequency spectroscopy \cite{Chin2004} and its momentum resolved extension \cite{Stewart2008}, Bragg spectroscopy \cite{Veeravalli2008} and lattice modulation spectroscopy \cite{Jordens2008,Greif2011}. Although these techniques have been very successful in characterizing strongly correlated systems, acquiring a sufficiently strong signal has always required creating multiple excitations. In addition, final-state interactions often complicate the interpretation of the obtained spectra.  

In this paper we propose a scheme for the measurement of momentum-resolved single-particle excitation spectra without final-state interactions, similarly to ARPES, using a quantum gas microscope. As illustrated in Fig.~\ref{fig1Setup} a) for a 1D spin system, the scheme involves modulating the tunneling amplitude between the chain and an initially empty detection system at a frequency $\omega_{\mathrm{shake}}$. By measuring the resulting transfer rate from the system to the probe, the spectral function $A(\omega,k)$ for a single hole inside the spin chain can be obtained. We present several ways how the momentum ($k$) can be resolved using the capabilities of quantum gas microscopy. It generalizes methods based on radio-frequency spectroscopy \cite{Chin2004,Dao2007,Torma2000,Chen2009a,Torma2016} and theory proposals to perform the equivalent of scanning tunneling microscopy on ultracold atoms \cite{Kollath2007}. 

To demonstrate our scheme, we consider variations of the $t-J$ model with isotropic and anisotropic spin interactions. The case of isotropic spin interactions has been realized experimentally as a limit of the 1D Fermi-Hubbard model at half-filling and strong coupling \cite{Mazurenko2016,Hilker2017}. Anistropic spin interactions can be realized with Rydberg dressing \cite{Zeiher2016,Zeiher2017}, using polar molecules \cite{Gorshkov2011tJ} or employing spin-dependent interactions \cite{Duan2003}. Theoretical calculations \cite{Szczepanski1990} have shown for both models that the shape of the spectral function can be understood from spin-charge separation. Here, we demonstrate that spinon and holon lines in the spectrum can be individually resolved at all energies in comparatively small systems of ten to twenty ultracold atoms at currently achievable temperatures. 

The ground state of the 1D $t-J$ model with isotropic spin interactions does not possess long-range order and is described by Luttinger liquid theory instead \cite{Giamarchi2003}. The spin-liquid nature of this ground state leads to an intriguing signature in the spectral function already for a single hole \cite{Szczepanski1990}: at low energies, most of the spectral weight is found for momenta $0 \leq k \leq \pi/2$, with lattice constant $a=1$, whereas between $\pi/2 < k \leq \pi$ the spectral weight is suppressed by several orders of magnitude, see Fig.~\ref{fig1Setup} b). This phenomenon is to some extent reminiscent of the Fermi arcs observed by ARPES in the pseudogap phase of cuprates \cite{Lee2006}. To explain the sharp reduction of spectral weight by a simple physical picture, we describe the Luttinger liquid ground state of the spin chain as a quantum spin liquid using slave-fermion mean field theory. In this formalism, the ground state of the Heisenberg chain with zero total magnetization can be understood as two identical, half-filled Fermi seas of spinons. As illustrated in Fig.~\ref{fig1Setup} c), the asymmetry in the spectrum $A(k,\omega)$ is easily understood by noting that the creation of a hole in an ARPES-type measurement corresponds to removing a spinon from one of the Fermi seas. In this work, we show that the asymmetry of the spectral function around $\pi/2$ in the $t-J$ model with isotropic spin couplings can be observed in experiments with ultracold atoms.

In contrast, the ground state of the anisotropic $t-J$ model with dominant Ising interactions between the spins is not a spin liquid, but possesses long-range N\'{e}el order. In this case the sub-lattice symmetry is spontaneously broken \cite{Giamarchi2003}, and the spectrum is approximately symmetric around $\pi/2$, i.e. $A(\pi/2 + k,\omega) \approx  A(\pi/2 - k,\omega)$. We extend the slave-fermion mean field theory to this regime and find that it correctly predicts the broken sub-lattice symmetry when $J_z/J_\perp$ is varied, where $J_\perp$ denotes the coupling strength in the $XY$-plane of the spins. Spinon excitations become gapped for $J_z>J_\perp$, and the mean field gap $\Delta$ is a non-analytic function of $J_z/J_\perp$ in agreement with exact Bethe ansatz calculations \cite{Sutherland}.

Our paper is organized as follows. In Sec.~\ref{secMeasurementScheme} we introduce the experimental scheme for measuring the spectral function using a quantum gas microscope. In Sec.~\ref{SecModels} we introduce two variations of the 1D $t-J$ model with isotropic and anisotropic spin couplings, for which we study the spectral function in Sec.~\ref{secSpectra1D}. We present results from exact numerical simulations which take into account effects of finite size and temperature. Two physical phenomena are discussed, which can be measured using our scheme: spin-charge separation for arbitrary energies (Sec.~\ref{subsecSpinChargeSep}) and the asymmetry of the spectrum, which is a signature of the Luttinger spin-liquid, for the case of isotropic spin couplings (Sec.~\ref{subsecSpectralSignaturesLuttSpinLiq}) and finite magnetization (Sec.~\ref{subsecSpinImbalance}). 
A theoretical analysis of our findings is provided in Sec.~\ref{secTheory}. In Sec.~\ref{secMF} we use a slave-fermion mean field theory to describe a spin chain and explain the asymmetry in the spectral function. Analytical results for the renormalization of a spin-less holon by collective spin excitations are discussed in Sec.~\ref{subsecHolonPolaron}.  Extensions to the measurement scheme are discussed in Sec.~\ref{secExtensions}. We close with a summary and by giving an outlook in Sec.~\ref{secOutlook}.

\section{Measuring spectral functions in a quantum gas microscope}\label{secMeasurementScheme}
In the following we outline our proposal to experimentally measure the spectral function of a single hole with simultaneous momentum and energy resolution in a quantum gas microscope.  The basic idea is to excite a single particle from a filled 1D system $S$ by lattice modulation into an adjacent 1D ``detection" system $S^{\mathrm{det}}$. The latter consists of empty sites and is offset in energy by $\Delta \gg t_y$ where $t_y$ is the bare tunneling amplitude between $S$ and $S^{\mathrm{det}}$, see Fig.~\ref{fig1Setup} a). The lattice modulation can be described by a perturbation term
\begin{equation}
\H_{\mathrm{pert}}(\tau)=\delta t_y \sin(\omega_{\mathrm{shake}} \tau)\hat{T}_y
\label{eq:Hpert}
\end{equation}
in the Hamiltonian. Here $\tau$ denotes time, $\delta t_y$ is the modulation amplitude of the hopping between $S$ and $S^{\mathrm{det}}$ described by the operator $\hat{T}_y$, and $\omega_{\mathrm{shake}}$ is the modulation frequency.

\subsection{Single-particle transfer}
A successful excitation transfers a single particle from $S$ to $S^{\mathrm{det}}$. As the modulation is only along the $y$-axis (i.e. perpendicular to the 1D system), the total momentum is conserved and the excitation couples simultaneously to all individual momenta $k$. This can be seen by rewriting the perturbation \eqref{eq:Hpert} in momentum space, 
\begin{equation}
\label{eq:pert}
\hat{T}_{y}  =  -\sum_{i, \sigma} \left(\hat{d}_{i, \sigma}^{\dagger} \hat{c}_{i, \sigma} + \mathrm{h.c.}\right)=- \sum_{k, \sigma} \left(\hat{d}_{k, \sigma}^{\dagger} \hat{c}_{k, \sigma} + \mathrm{h.c.}\right).
\end{equation}
Here $\hat{c}_{i(k), \sigma}$ denotes the annihilation operator at site $i$ (momentum $k$) in $S$ and $\hat{d}_{i(k), \sigma}^{\dagger}$ denotes the respective creation operator in $S^{\mathrm{det}}$. The spin-index is $\sigma=\uparrow,\downarrow$. The energy change of the system with one hole as compared to the initial state without a hole is $\hbar \omega = E^{N-1}-E^{N}$. For a lattice modulation frequency $\omega_{\mathrm{shake}}$ this is determined by energy conservation,
\begin{equation}
\hbar \omega = \hbar\omega_{\mathrm{shake}} - E^s(k) - \Delta,
\end{equation}
where $\Delta$ is the energy offset and $E^s(k) = -2t \cos(ka)$ is the energy of the particle in the detection system, with $t$ the hopping amplitude of the particle in $S^{\mathrm{det}}$.
As explained in Sec.~\ref{subSecMomentum}, a subsequent momentum-space mapping technique of the single particle in $S^{\mathrm{det}}$ allows one to determine the momentum $k$ of the transferred atom. Thus, both full momentum and energy resolution are achieved.

By measuring the final position of the transferred atom and repeating the same measurement for various lattice modulation times, the excitation rate $\Gamma(k, \omega)$ can be determined. This rate quantifies the probability for creating a hole with momentum $k$ and energy $\hbar \omega$ in $S$, normalized by the modulation time. Up to constant pre-factors, it is identical to the hole spectral function,
\begin{equation}
 \Gamma(k, \omega)= \frac{2\pi}{\hbar} |\delta t_y|^2 A(k, \omega),
\end{equation}
as obtained by Fermi's golden rule. 

The spectral function of the hole $A(k,\omega)$ is defined as
\begin{multline}
A(k,\omega)= \frac{1}{Z_0} \sum_{n,m} \sum_\sigma e^{-\beta E_n^{N}} |\bra{\psi_m^{N-1}} \hat{c}_{k,\sigma} \ket{\psi_n^{N}} |^2 \\
\times \delta (\hbar \omega - E_m^{N-1} + E_n^{N}),
\label{eq:spectralfunction}
\end{multline}
with $\ket{\psi_n^N}$, $E_n^N$ denoting the eigenstates and -energies of the system $S$ with $N$ particles. Furthermore, $\beta=1/k_B T$ is the inverse temperature and $Z_0 = \sum_n e^{-\beta E_n^{N}}$ denotes the partition function before the perturbation Eq.~\eqref{eq:Hpert} is switched on. 

For small system sizes it is important to choose a sufficiently small excitation amplitude $\delta t_y/t_y$, such that at most a single particle is transferred, in order to avoid multiple excitations as well as final state interactions. The latter can also be avoided by implementing a spin-changing Raman transfer to a non-interacting spin state instead of a lattice modulation. For large systems, we expect multiple excitations to not alter the spectral function as long as the average fraction of excited particles remains sufficiently small.

\subsection{Momentum Resolution}
\label{subSecMomentum}
A crucial step for measuring the spectral function is the momentum detection in the probe system $S^{\mathrm{det}}$. This can be achieved by combining the capabilities of a quantum gas microscope with a digitial micromirror device (DMD), which gives control over the optical potential of the atoms on a site-resolved level. This precise control has already been demonstrated with bosonic and fermionic atoms with single-site resolution \cite{Zupancic2016, Mazurenko2016}. By illuminating the DMD with blue-detuned light, a box-like potential with hard walls at the two ends of the 1D systems can be created. This limits the size of both systems $S$ and $S^{\mathrm{det}}$ to $L$ sites. By adding a parabolic potential, any harmonic confinement in the 1D system caused by the underlying Gaussian beam shape of the lattice beams can additionally be cancelled over the region of interest. The box geometry ensures that the absolute value of the momentum $|k|$ of the transferred particle remains unchanged after the action of the perturbation $\H_{\mathrm{pert}}(\tau)$, while still confining the particle within $S^{\mathrm{det}}$.

The perturbation is followed by a bandmapping step, which converts momentum space into position space. Subsequent site-resolved imaging then allows one to reconstruct the particle's momentum. We now discuss three possibilities how such a mapping procedure can be implemented and give an estimate for the achievable momentum resolution in typical experimental setups. The momentum resolution $\kappa$ is quantified by the inverse number $N_k$ of different momentum states in the lowest band with $|k| < \pi/a$ that are detectable,
\begin{equation}
\kappa = 1 / N_k.
\end{equation}

\begin{figure}[b!]
\centering
\epsfig{file=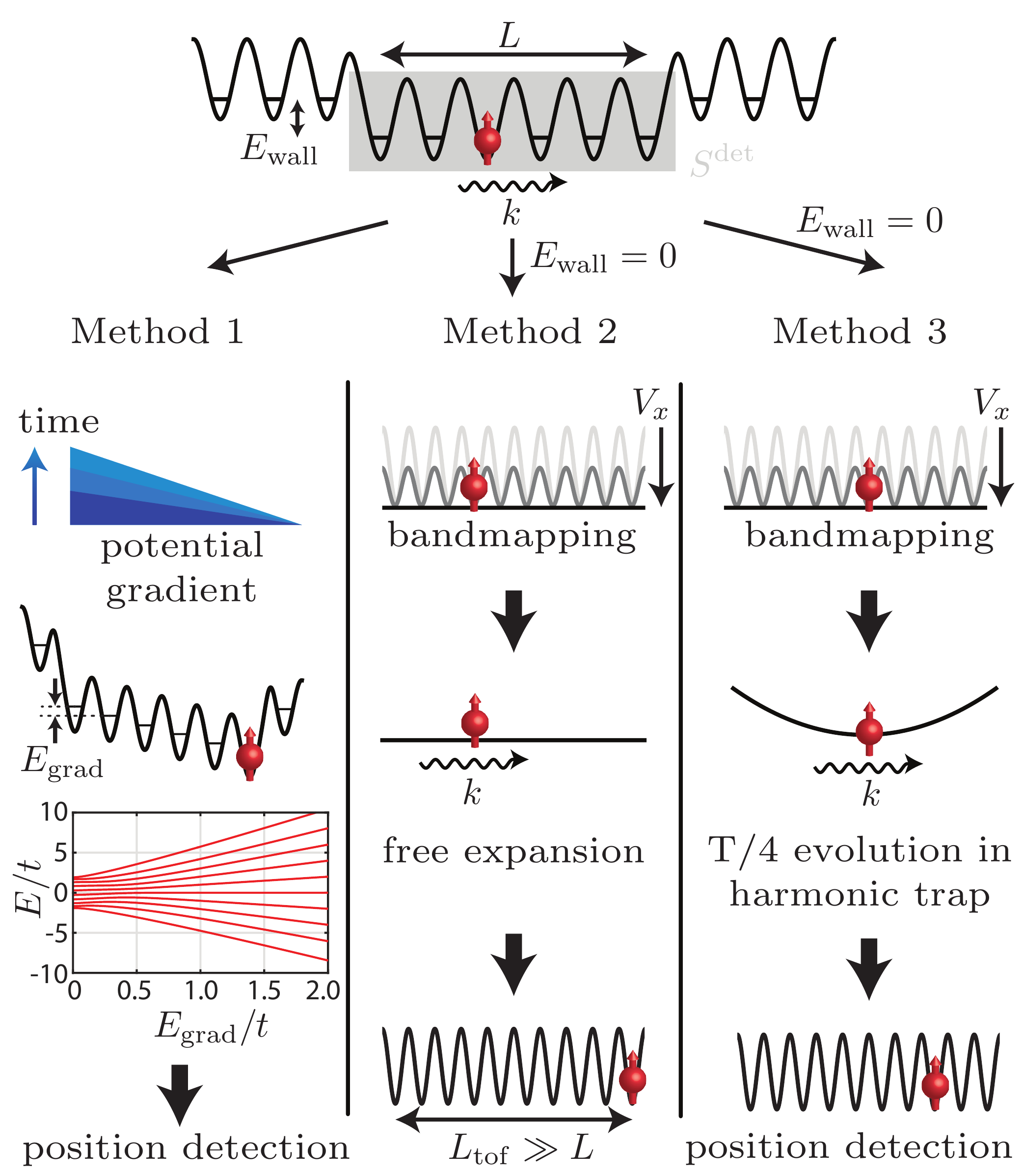, width=\columnwidth}
\caption{\textbf{Measuring the momentum of the excitation.} The momentum of the hole-excitation in $S$ is measured from the momentum of the excited particle in $S^{\mathrm{det}}$, which has a finite size of $L$ sites determined by the energy offset $E_{\mathrm{wall}}$ at the edges. $E_{\mathrm{wall}}$ is chosen to be larger than all relevant energy scales in $S$. The three methods discussed in the main text are illustrated. The exemplary diagram shown in the first column illustrates that the eigenenergies in $S^{\mathrm{det}}$ are smoothly connected when introducing a potential gradient $E_{\mathrm{grad}}$ (here $L=10$). For $E_{\mathrm{grad}}\gg t$ the eigenstates are localized on individual lattice sites. For the first method we keep $E_{\mathrm{wall}}$ unchanged, whereas we set $E_{\mathrm{wall}}=0$ for the other two methods before the bandmapping. }
\label{fig2Momentum}
\end{figure}

\subsubsection{Wannier-Stark mapping}
The first method for mapping momentum space into position space is to smoothly introduce a potential gradient along the $x$-direction, which causes an energy shift of $E_{\mathrm{grad}}$ per lattice site. Such a potential gradient can be implemented for example by applying a magnetic field gradient exploiting the atomic Zeeman shift or by using the DMD. In the limit of a vanishing gradient $E_{\mathrm{grad}}\ll t$ the single-particle energy eigenstates in $S^{\mathrm{det}}$ are quasi-momentum states $E^s(k_n)$ with discrete momenta $k_n=n \pi/L$ owing to the finite size of the box. For very large gradients $E_{\mathrm{grad}}\gg t$ the eigenstates are Wannier-Stark states localized on single lattice sites and separated in energy by $E_{\mathrm{grad}}$. As shown in the left column of Fig.~\ref{fig2Momentum}, these eigenstates are smoothly connected for an increasing potential gradient $E_{\mathrm{grad}}$. 

The momentum resolution of this method is determined by the initial number of lattice sites in $S^{\mathrm{det}}$ and is given by $1/L$. Adiabatic mapping requires the gradient ramp time to be much slower than the smallest energy splitting $\delta E^s$, which in this case is given by the energy spacing between adjacent quasi-momentum states at $E_{\mathrm{grad}}=0$. The finite lifetime of atomic quantum gases sets an upper limit for the gradient ramp time and thus a lower limit to $\delta E^s$. This limits the maximum box size and hence the momentum resolution of this method. Experimentally, ramp timescales of hundreds of tunneling times are routinely used in lattice loading protocols, corresponding to an energy of about $0.01t$ \cite{Parsons2016}. Assuming a tenfold slower gradient ramp time to ensure adiabaticity ($\delta E^s=0.1t$), we find $L=20$. This demonstrates that already this simple scheme gives a very good momentum resolution of about $\kappa \approx 1/20$. Furthermore, the ramp velocity can be increased at later times in the protocol, since the energy spacings become larger with growing $E_{\mathrm{grad}}$, thus enhancing the momentum resolution.  

\subsubsection{Time-of-flight mapping}
An alternative method of determining the momentum is to perform a time-of-flight expansion along the $x$-direction in $S^{\mathrm{det}}$ after exciting the single particle. This can be done by suddenly turning off the DMD light which creates the box potential and applying a bandmapping of the lattice in the $x$-direction, see Eq.~\ref{eq:optical_potential}. This maps quasi-momentum states into momentum states \cite{Greiner2001} of $S^{\mathrm{det}}$. Ballistic expansion of the single particle along the $x$-direction for a duration of $\tau_{\mathrm{tof}}$ and subsequent detection of the displaced atomic position $x_{\mathrm{tof}}$ using the quantum gas microscope then allows one to determine the atomic momentum via $k=\pi m \lambda_x^2 x_{\mathrm{tof}}/(2h\tau_{\mathrm{tof}})$, where $m$ is the atomic mass, $h$ is the Planck constant and $x_{\mathrm{tof}}$ and $k$ are normalized to the lattice spacing. 

This procedure requires a sufficiently long time-of-flight expansion such that the initial system size is negligible, i.e. $2h\tau_{\mathrm{tof}}/(m\lambda_x^2)\gg L $. During the detection procedure the lattice depths along the $y$-direction remain unchanged to ensure that the particle remains trapped inside the 1D tube. The largest achievable value of $\tau_{\mathrm{tof}}$ is determined by the largest spatial separation $L_{\mathrm{tof}}$ under the microscope where site-resolved imaging can still be reliably performed. As the particles are initially located in a box of $L$ sites, there are also $L$ momentum states. After free expansion to a size of $L_{\mathrm{tof}}$, a particle initially in a momentum state will then be detected within a spatial region that approaches $L_{\mathrm{tof}}/L$ sites for long time-of-flight times. Corrections due to a finite time-of-flight are therefore negligible if this size exceeds the initial system size $L$. From this we obtain an upper bound for the initial system size of $L=\sqrt{L_{\mathrm{tof}}}$. In addition, clean mapping requires a flat system along the $x$-direction after the bandmapping. The harmonic confinement along that direction caused by the $y$-lattice beams can be cancelled by a blue detuned anti-confinement beam created by a DMD or Gaussian beam with a suitable beam waist.  

In bosonic quantum gas microscopy a related variant of the proposed technique has already been implemented, where atoms in a small system of a few sites were expanded in 1D tubes to a width of about $L_{\mathrm{tof}}=100$ sites and successfully detected with single-site resolution \cite{Islam2015}. For these parameters we estimate a momentum resolution of $\kappa \approx 1/10$ for our scheme. 

\subsubsection{$T/4$-mapping}
A third technique for mapping momentum-space into real space that does not rely on a long expansion distance is based on a quarter period rotation in phase-space in the presence of a harmonic trap \cite{Murthy2014}. After suddenly introducing an underlying harmonic confinement with period $T$ into the probe system $S^{\mathrm{det}}$, the real-space distribution after a time evolution of $T/4$ will precisely correspond to the initial momentum distribution of the transferred particle (and vice versa). 

To achieve this, we propose to first suddenly turn off the DMD light for the box potential and apply a bandmapping of the $V_x$ lattice, as before. Then a strong harmonic confinement can be introduced by rapidly increasing the lattice depth along the $y$-direction. This leads to an increased harmonic confinement along the $x$-direction owing to the Gaussian beam shape of the laser beam. Alternatively, a DMD with red-detuned light could be used. After letting the single particle in $S^{\mathrm{det}}$ evolve for a quarter period, its position can be measured with the quantum gas microscope. The advantage of this method compared to the previous one is that it does not require imaging over large distances for good momentum resolution. By adjusting the frequency $\omega$ of the strong harmonic trap, the largest displacement of the single particle relative to the center of the box can be controlled. It can be chosen to be comparable to the initial system size $L$. Assuming a maximum imaging width of $100$ sites (as before), this method would allow a momentum resolution of about $\kappa \approx 1/100$. 

Current typical sizes of fermionic lattice systems at low temperatures with single-site resolution are on the order of $10$ sites \cite{Mazurenko2016, Hilker2017}. The highest desirable momentum resolution is therefore $\kappa = 1/10$, which would be provided by all three proposed  methods. In the future, when larger system sizes become available experimentally, the Wannier-Stark mapping and the quarter-period rotation scheme promise the highest momentum resolution.

\section{The Models}
\label{SecModels}

In this section, we introduce the two models on which our theoretical calculations are performed. Both Hamiltonians are closely related to the $t-J$ model. Note, however, that our scheme for measuring the spectral function is not specific to these models.

\subsection{The $t-J^*$ model}

The 1D Fermi-Hubbard model is described by the Hamiltonian
\begin{equation}
\H_{\rm FH} = -t \sum_{\ij, \sigma} \cd_{i,\sigma} \c_{j,\sigma} + U \sum_{j} \hat{n}_{j,\uparrow} \hat{n}_{j,\downarrow}.
\label{eq:fhmodel}
\end{equation}
Here, $\cd_{j,\sigma}$ creates a fermion with spin $\sigma$ on site $j$ and $\n_{j,\sigma} = \cd_{j,\sigma} \c_{j,\sigma}$ denotes the density operator of fermions with spin $\sigma$. The local Hubbard interaction is given by $U$ and fermions are hopping with rate $t$ between neighboring sites $\ij$. 

In the large-$U$ limit and below half filling the Fermi-Hubbard Hamiltonian \eqref{eq:fhmodel} can be mapped to the $t-J^*$ model. Up to order $\mathcal{O}(t^2/U)$ the exact representation is
\begin{widetext}
\begin{equation}
\H_{t-J^*} =  \mathcal{P} \biggl[ -t \sum_{\ij, \sigma} \cd_{i,\sigma} \c_{j,\sigma} + J \sum_{j} \left(\hat{\mathbf{S}}_{j+1} \cdot \hat{\mathbf{S}}_j - \frac{ \hat{n}_{j+1} \hat{n}_j }{4}\right)  - \frac{J}{8}  \sum_{\langle i,j,r \rangle, \sigma}^{i \neq r}  \biggl( \cd_{i,\sigma} \c_{r,\sigma} \n_j  - \sum_{\sigma', \tau, \tau'} \cd_{i,\sigma} \vec{\sigma}_{\sigma,\sigma'} \c_{r,\sigma'} \cdot  \cd_{j,\tau} \vec{\sigma}_{\tau, \tau'} \c_{j,\tau'} \biggr) \biggr] \mathcal{P},
\label{eq:tjmodel}
\end{equation}
\end{widetext}
see e.g. Ref.~\cite{Auerbach1998}. Here, $\mathcal{P}$ denotes the projection operator on the subspace without double occupancy, and $\langle i,j,r \rangle$ is a sequence of neighboring sites. The operator $\cd_{j,\sigma}$ creates a fermion with spin $\sigma$ on site $j$ and $\n_{j,\sigma} = \cd_{j,\sigma} \c_{j,\sigma}$ denotes the density operator of fermions with spin $\sigma$. The spin operators are defined by $\hat{\vec{S}}_j = \frac{1}{2} \sum_{\sigma,\sigma'} \cd_{j,\sigma} \vec{\sigma}_{\sigma,\sigma'} \c_{j,\sigma'}$, where $\vec{\sigma}$ denotes a vector of Pauli matrices. The first term in Eq.~\eqref{eq:tjmodel} describes tunneling of holes with amplitude $t$. The second term corresponds to spin-exchange interactions of Heisenberg type, with anti-ferromagnetic coupling constant $J=4 t^2/U$. For a single hole, the term $\hat{n}_{j+1}\hat{n}_j$ leads to a constant shift in energy, which we will not include in the analysis in the following sections. Together these first two terms define the $t-J$ model. It is extended to the $t-J^*$ model by including the last term, which describes next-nearest neighbor tunneling of holes correlated with spin-exchange interactions. 

We discuss in more detail in Appendix \ref{Appdx1exptjmodel} how the measurement scheme for the spectral function can be implemented for the $t-J^*$ model using ultracold fermions in optical lattices.

\subsection{The $t-$XXZ model}

The $t$-XXZ model is described by the Hamiltonian
\begin{multline}
\H_{t-{\rm XXZ}} =  \mathcal{P} \biggl[ -t \sum_{\ij, \sigma} \cd_{i,\sigma} \c_{j,\sigma} + J_z \sum_{j} \hat{S}_{j+1}^z \hat{S}_j^z \\
+ \frac{J_\perp}{2} \sum_{j}\left( \hat{S}_{j+1}^+\hat{S}_j^- + \hc \right)  \biggr] \mathcal{P},
\label{eq:tjzmodel}
\end{multline}
with the same terminology as introduced above. Hamiltonians closely related to Eq.~\eqref{eq:tjzmodel} can be realized in a quantum gas microscope using polar molecules \cite{Gorshkov2011tJ}, Rydberg dressing \cite{Zeiher2016,Zeiher2017} or by spin-dependent interactions \cite{Duan2003}. In this case, there is no next-nearest neighbor hole hopping term. Furthermore, anisotropic spin coupling constants can also be realized with spin-dependent lattices \cite{Brown2015}.

\section{Spectra of holes in the 1D antiferromagnetic spin chains}
\label{secSpectra1D}

\begin{figure*}
\centering
\epsfig{file=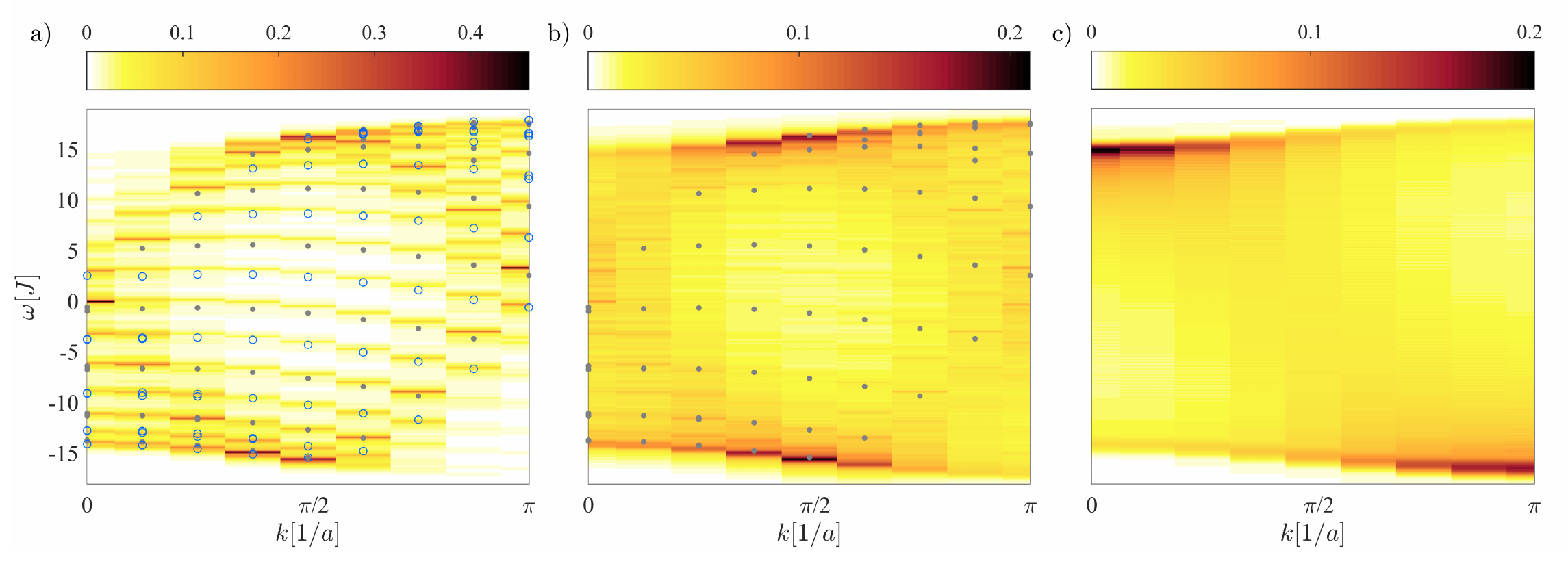, width=\textwidth}
\caption{\textbf{Temperature dependence of the spectral function.} We consider the $t-J^*$ model with periodic boundary conditions and $L=16$ sites for $t=8J$ at temperatures a) $T=0.4 J $, b) $T=0.75J$ and c) $T=5J$. In a) and b), a comparison with peak positions expected from the spectral building principle due to holon and spinon dispersions, Eq.~\eqref{eq:dispersion}, is provided (gray dots). Additionally, in a) open blue circles denote peaks expected from the spectral building principal due to low energy excitations in the spin chain relevant at finite temperatures, see text.}
\label{fig3Temperature}
\end{figure*}

In the following we present numerical results for the spectral function of a single hole in a one-dimensional, antiferromagnetic spin chain, see Fig.~\ref{fig3Temperature}. Similar results for periodic boundary conditions and at zero temperatures have been obtained, e.g., in Refs.~\citep{Bannister2000,Szczepanski1990,Eder1997,Kim1997}. Here we generalize those studies to systems with open boundary conditions, finite temperatures and spin imbalance. Several ARPES measurements have been performed in quasi-one dimensional materials, see e.g. Refs.~\cite{Kim1996,Kim1997}, and direct signatures of independent spinon and holon branches have been found at low energies \cite{Kim2006}.

The spectral function as defined in Eq. \eqref{eq:spectralfunction} is related to the Green's function of the hole via $A(k,\omega) = - (1/\pi) \text{Im} G(k,\omega)$ and can be calculated using standard Lanczos techniques. The $\delta$-peaks obtained by this means are slightly broadened to end up with a smoother spectral function.

\begin{figure}
\centering
\epsfig{file=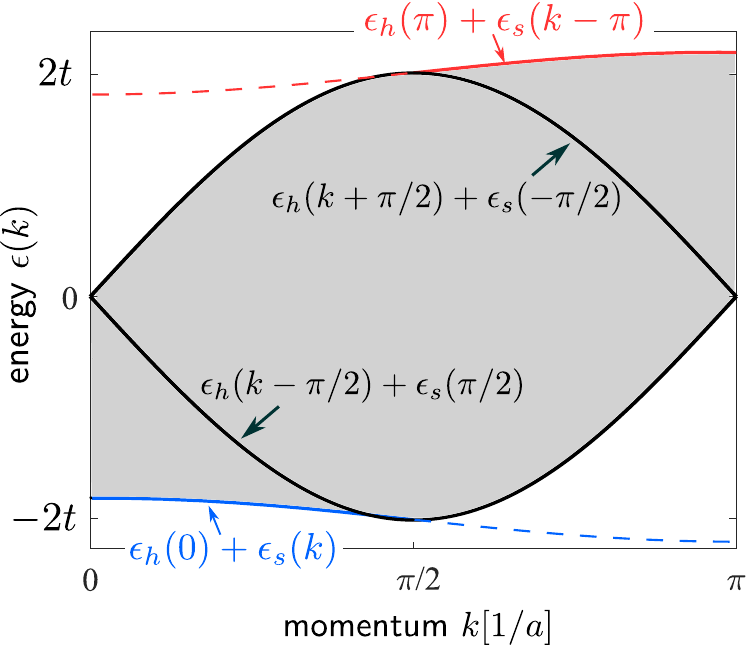, width=0.4\textwidth} $\qquad$
\caption{\textbf{Spectral building principle.} The combined dispersion relation of spinon and holon, Eq.~\eqref{eq:dispersion}, can be constructed by fixing the spinon or holon momentum, $k_s$ or $k_h$, and varying the other momentum, respectively. Because the spinon dispersion is only defined from $k_s=-\pi/2 ... \pi/2$ for zero temperature the spectrum is strongly asymmetric around $k=\pi/2$ at low and high energies (blue and red boundaries). In this case spectral weight can only be found in the shaded areas. The dashed blue and red lines indicate the lower and upper boundaries obtained when the spinon dispersion is extended to $k_s = -\pi ... \pi$ assuming $\epsilon_s(k_s) = \cos(k_s) J \pi / 2$. This case is relevant for high temperatures.}
\label{specConstr}
\end{figure}

\subsection{Spin-charge separation at arbitrary energies}
\label{subsecSpinChargeSep}

In this subsection, we consider the $t-J^*$ model, see Eq.~\eqref{eq:tjmodel}. Remarkably, a single hole moving with hopping amplitude $t$ in an antiferromagnetic spin chain with coupling $ J \ll t$ can be approximated by an almost free hole that is only weakly coupled to the spin chain. This constitutes a microscopic manifestation of spin-charge separation in 1D systems.

In order to understand the main features in the spectral function shown in Fig.~\ref{fig3Temperature}, it is important to distinguish between the \textit{holon} and the \textit{spinon}. The holon is the charge excitation, whereas the spinon is the spin $1/2$ excitation associated with the creation of a hole. As a consequence of spin-charge separation in 1D, the holon propagates on a timescale set by the hopping amplitude $t$ and is largely decoupled from the dynamics of the spinon, which moves on a timescale set by the exchange energy $J$. We can thus apply the semi-phenomenological spectral building principle, see \citep{Eder1997,Bannister2000} and Fig.~\ref{specConstr}, where the spinon and the holon are treated as independent particles, to determine where spectral weight is expected. A microscopic explanation for the $t-J^*$ model based on a slave-fermion mean field theory is provided in Sec.~\ref{secTheory}. Corrections to this picture due to the coupling of the holon to collective spin excitations will be derived in Sec.~\ref{subsecHolonPolaron}. 

The dispersion relation of a free spinon,
\begin{equation}
\epsilon_s(k_s) = J\frac{\pi}{2} |\cos( k_s)|, \qquad -\pi/2 \leq k_s \leq \pi/2,
\label{eq:spinon_dispersion}
\end{equation}
is a result of exact Bethe ansatz calculations for the isotropic spin chain \cite{Giamarchi2003}. 
The holon dispersion 
\begin{equation}
\epsilon_h(k_h) = - 2 t \cos ( k_h ) - \frac{1}{4} J \cos ( 2k_h )
\label{eq:holon_dispersion}
\end{equation}
corresponds to the one of a free particle and can be derived from the $t-J^*$ Hamiltonian itself. The second term stems from the next nearest neighbor hopping of the hole, see Eq.~\eqref{eq:tjmodel}.

As a consequence of conservation of energy $\epsilon$, the free spinon and holon dispersions can be combined to
\begin{align}
\begin{split}
\epsilon(k) &= \epsilon_h(k_h) + \epsilon_s(k_s) 
\\
&= - 2 t \cos (k_h)  - \frac{1}{4} J \cos (2k_h) + J \frac{\pi}{2} |\cos (k_s)|.
\end{split}
\label{eq:dispersion}
\end{align}
By momentum conservation it holds that $k=k_s + k_h$. Thus we can set $k_s = k - k_h$ in Eq.~\eqref{eq:dispersion} and for a given $k$ regard the holon momentum $k_h$ as a free parameter. The energy $\epsilon(k) = E_m^{N-1} - E_n^N$ in Eq.~\eqref{eq:dispersion} enters the $\delta$-function in the Lehmann representation of $A(k,\omega)$, Eq.~\eqref{eq:spectralfunction}, and allows one to predict the positions of peaks in the spectral function.

In an infinite system, Eq.~\eqref{eq:dispersion} can be used to determine the region in $k-\omega$ space where spectral weight exists at zero temperature. As shown in Fig.~\ref{specConstr}, the boundaries of this region are determined by the spinon and holon dispersions, respectively. In a finite system with $L$ sites, the quantization of the holon momentum leads to $L$ distinct lines, as can be seen in Figs.~\ref{fig1Setup} b) and \ref{figIsing} a). This indicates that the spectrum can be well described by non-interacting spinons and holons: each quantized holon momentum $k_h$ can be associated with a branch in the spectrum obtained by changing the spinon momentum $k_s$ and keeping $k_h$ fixed.

In Fig.~\ref{fig3Temperature}, we investigate the influence of finite temperatures on the spectral function $A(k,\omega)$ of a single hole for the $t-J^*$ model at half filling. We only plot the spectral function for momenta $k$ with $0 \leq k \leq \pi$, since the spectrum is symmetric around $k=0$. Gray dots denote the combined dispersion relations of holon and spinon, Eq.~\eqref{eq:dispersion}, where the spectral building principle predicts peaks in the spectral function. 
We have included shifts in the spinon and holon momentum due to their different quantization conditions and a twisted periodic boundary effect, which are explained in detail in Appendix~\ref{AppdxShifts}.

At sufficiently low temperatures, the peaks expected from the spectral building principle coincide with the peaks of the spectral function calculated with the help of Lanczos techniques, Fig.~\ref{fig3Temperature}a). This indicates that spinon and holon can indeed be treated separately. Moreover, spin charge separation is not restricted to the low frequency part, but can be observed across the whole spectrum. Thermal fluctuations in energy and momentum lead to a broadening of the peaks predicted by the spectral building principle. 

In the case of periodic boundary conditions, additional peaks, marked by blue circles in Fig.~\ref{fig3Temperature} a), appear between the lines found at zero temperature. This can be understood by considering the effect of thermal excitations in the spin chain. The lowest-energy states, which are most relevant at small temperatures, carry momentum close to zero and close to $\pi$. As explained in detail in Appendix \ref{AppdxShifts}, an excitation in the spin chain with momentum $\pi$ introduces a twisted periodic boundary effect for the holon and thereby gives rise to the additional lines observed in the spectrum.

For increasing temperatures $T \gtrsim J$, Fig.~\ref{fig3Temperature}b) and c), the spectral building principle starts to break down. While the shape of the lower edge still corresponds to the spinon dispersion \cite{Hayn2000}, more and more low energy excitations start to appear for $\pi/2 \leq k \leq \pi$. Furthermore, as the temperature approaches and exceeds $J$, the distinct lines are replaced by a continuum, demonstrating that there is no longer a single well-defined spinon. 
The comparison of Fig.~\ref{fig3Temperature}b) and c) shows a shift of spectral weight from $k=\pi/2$ at the upper and lower boundary of the spectrum to $k=0$ and $k=\pi$, respectively. At high temperatures $T\gg J$, the distribution of spectral weight is essentially determined by the density of states.

\subsection{Effect of anisotropy on the spectral function}
\label{subsecSpectralSignaturesLuttSpinLiq}

\begin{figure}[t!]
\centering
\includegraphics[width=0.5\textwidth]{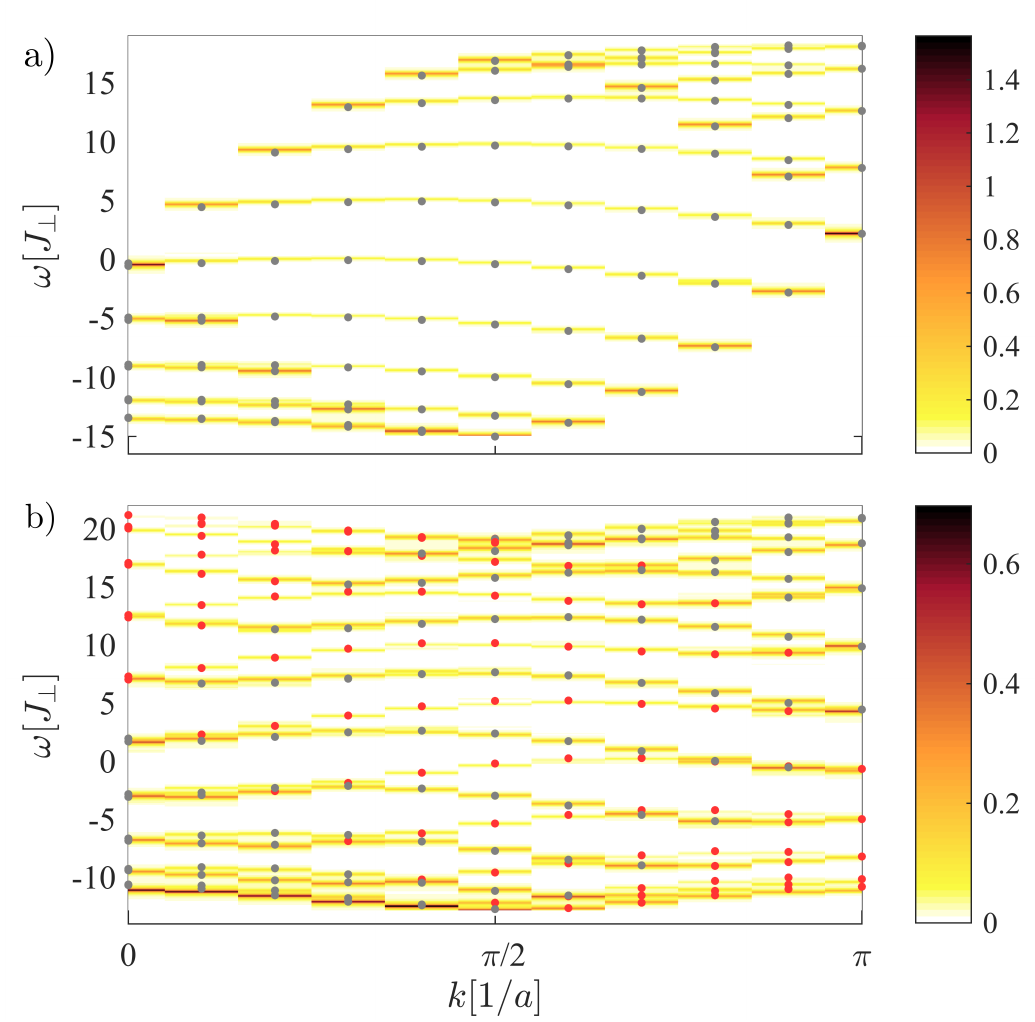}
\caption{\textbf{Zero temperature spectra} for a system with 20 sites, periodic boundary conditions, described by the $t$-XXZ Hamiltonian, Eq~\eqref{eq:tjzmodel}, with hopping $t=8J_\perp$ and a) $J_z=J_\perp$, b) $J_z = 4 J_\perp$. Gray dots correspond to spinon and holon dispersion, see Eq. \eqref{eq:dispersion} and Eq. \eqref{eq:dispersion_ising}, respectively, with the spinon momentum restricted to half the Brillouin zone, $ -\pi/2 \leq k_s \leq \pi/2$. Red dots in b) denote spinon momenta in the remaining half of the Brillouin zone, where no spectral weight appears in the isotropic case $J_z=J_\perp$.}
\label{figIsing}
\end{figure}

We now consider the $t$-XXZ model as introduced in Eq. \eqref{eq:tjzmodel} which is characterized by an anisotropy in the coupling constants of the spins.
In Fig.~\ref{figIsing}, the spectral function for this model is shown for a) $J_z = J_\perp$ and b) $J_z = 4 J_\perp$. The effects of spin-charge separation discussed for the isotropic Heisenberg spin chain in Sec. \ref{subsecSpinChargeSep} appear here as well and the spectral building principle can be applied with a modified spinon dispersion relation. First order perturbation theory in $J_\perp/J_z$ leads to the spinon dispersion $\epsilon_s^{\rm Is}(k_s)= J_\perp \cos (2 k_s) + J_\perp+J_z$, such that for $J_z>J_\perp$
\begin{align}
\epsilon(k) = - 2 t \cos ( k_h ) + J_\perp \cos \l 2(k-k_h) \r + J_\perp+J_z.
\label{eq:dispersion_ising}
\end{align}
In Fig.~\ref{figIsing}, gray dots correspond to $\epsilon(k)$ from Eq.~\eqref{eq:dispersion} for $J_\perp=J_z$ and Eq.~\eqref{eq:dispersion_ising} for $J_z>J_\perp$, respectively, with $|k-k_h| \leq \pi/2$. In Fig.~\ref{figIsing} b), red dots denote $\epsilon(k)$ for values $|k-k_h| > \pi/2$ where no spectral weight appears in the isotropic case. 

In comparison to Fig.~\ref{figIsing} a), where spectral weight appears only for spinon momenta $|k-k_h| \leq \pi/2$ in one half of the Brillouin zone, the spectrum for an anisotropic spin chain features an almost symmetrical distribution of spectral weight around $|k-k_h| = \pi/2$.

\begin{figure}[t!]
\centering
\epsfig{file=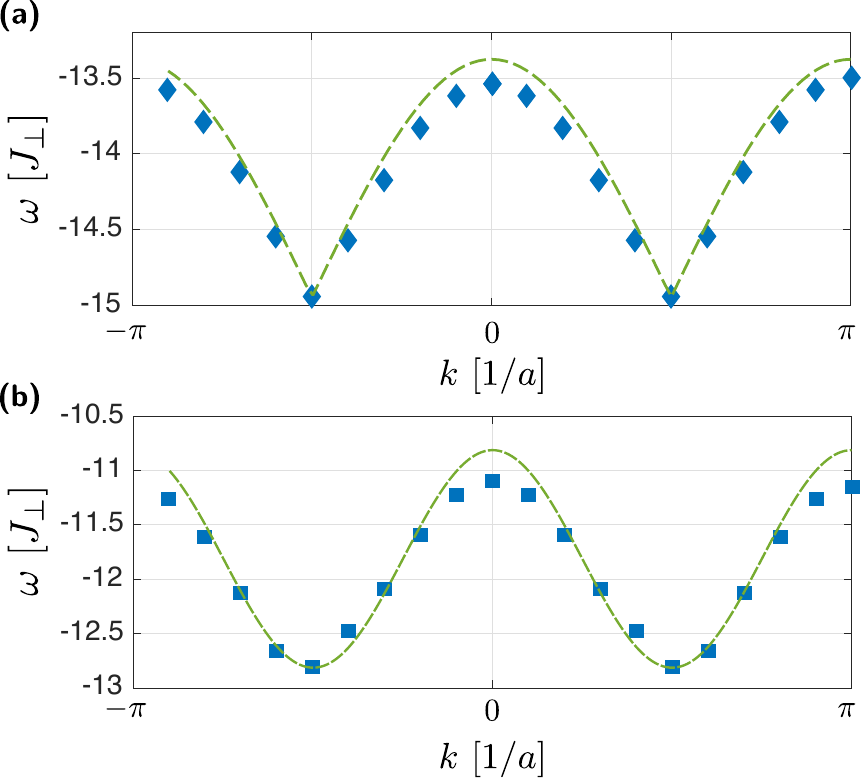, width=0.45\textwidth}
\caption{\textbf{The ground state energy as a function of the total momentum} $k$ is shown for a single hole in a spin chain (full symbols). We used the same parameters as in Fig.~\ref{figIsing} a) and b) respectively. The dashed line corresponds to the free spinon dispersion a) in the Heisenberg model with $J_z=J_\perp$ and b) in the XXZ model with $J_z=4 J_\perp$. }
\label{fig3SCseparation}
\end{figure}

In principle, there can be several reasons why in the isotropic case no spectral weight is observed at low energies for $k > \pi/2$. The most obvious one is, that there are no eigenstates for the corresponding energies and momenta. As we demonstrate in Fig.~\ref{fig3SCseparation}, this is not the case here: We calculate the ground state energy of the spin chain, doped with a single hole, as a function of the total momentum $k$ with exact diagonalization. For $0 \leq |k| < \pi/2$, the exact ground state energy closely follows the spinon dispersion for arbitrary $J_z/J_\perp$. Both in the isotropic and anisotropic case, see Fig.~\ref{fig3SCseparation} a) and b), there are low-energy eigenstates for all $k$ and to a good approximation their energies are symmetric around $k=\pi/2$. We conclude that for the isotropic spin chain there exist low-energy eigenstates for $k > \pi/2$, but their spectral weight is strongly suppressed.

Another possible reason for the strong suppression of the spectral weight could be a selection rule \cite{Szczepanski1990}, caused for example by the $SU(2)$ symmetry of the system at the isotropic point $J_\perp = J_z$. Because the ground state of the Heisenberg model is a singlet, only states with total spin $S=1/2$ and one hole can have finite weight in the spectrum at zero temperature. It has been found by exact numerical simulations in Ref.~\cite{Szczepanski1990} that this selection rule indeed applies for the ground state of the spin chain with one hole at momenta $|k| >\pi/2$, which has $S=3/2$. However, at only slightly higher energies of the order of $J$, we have numerically found eigenstates with $S=1/2$, for which the selection rule does not apply. In fact, these states give rise to a non-zero spectral weight at low energies for $|k| > \pi/2$. Since it is suppressed by about three orders of magnitude compared to the spectral weight observed at the same energies for $|k| < \pi/2$, it is not noticible in Fig.~\ref{figIsing} a). In contrast to what has been suggested in Ref.~\cite{Szczepanski1990}, a selection rule seems not to be sufficient to explain the asymmetry of the spectral weight observed for a hole created in a Heisenberg chain.

Above, we have ruled out the simplest two explanations why the spectral weight at low energies is almost completely restricted to one half of the Brillouin zone in the isotropic case, $J_z=J_\perp$. This effect hints at a more fundamental structure in the ground state wavefunction of the one-dimensional Heisenberg antiferromagnet. In contrast to the Ising case, the Heisenberg spin chain has singlet character and can be understood as a resonating valence-bond state \cite{Auerbach1998}. So, it is interesting to ask whether the valence-bond character of the ground state wavefunction is sufficient to explain the sharp drop of the spectral weight when the spinon momentum crosses $k=\pi/2$. We have checked that this is not the case, by calculating the spectral function for a hole created inside a spin chain with Majumdar-Gosh couplings \cite{Majumdar1969a,Majumdar1969}, see Appendix \ref{Appdx1MG}. The ground state of this model is a valence bond solid. While the spectral weight is asymmetric around $k=\pi/2$ in this case, it smoothly drops as the spinon momentum $\pi/2$ is traversed.

We argue instead that the sharp decrease of the spectral weight for a single hole in the Heisenberg chain can be understood as a direct signature for the presence of a Fermi sea of spinons, see Fig.~\ref{fig1Setup} c). This is characteristic for a quantum spin-liquid \cite{Wen2004}. In Ref.~\cite{Kim1997} it has been suggested that the Fermi sea is formed by Jordan-Wigner fermions, which can be introduced by fermionizing the spins. However, at the isotropic point $J_z=J_\perp$ these Jordan-Wigner fermions are strongly interacting, and the non-interacting Fermi sea is not a good approximation. Instead, slave fermions can be introduced as in the usual mean field description of quantum spin liquids \cite{Wen2004}. They are weakly interacting at the isotropic point and form two half-filled spinon Fermi seas. These arguments are supported by slave particle mean field calculations. In Sec.~\ref{secMF} of this paper we present a mean field theory for a single hole and arbitrary values of the anisotropy $J_\perp/J_z$. A related work on slave-particle mean field descriptions of one-dimensional spin chains has been presented in Ref.~\cite{Weng1995}. In our paper we utilize the slave-fermion theory to analyze the spectral function.

\subsection{Spectral function of spin-imbalanced systems}
\label{subsecSpinImbalance}

In the slave-particle mean field picture, the slave fermions form two spinon Fermi seas. Therefore we expect to see two different Fermi momenta when the system is spin-imbalanced. Our scheme to measure the spectral function in experiments with cold atoms is particularly well suited to access the spectral function of a single hole in a system with finite magnetization. Moreover, by detecting the spin of the removed particle~\cite{Boll2016}, the spin-resolved version of the spectral function can be measured.

\begin{figure}[t!]
\centering
\epsfig{file=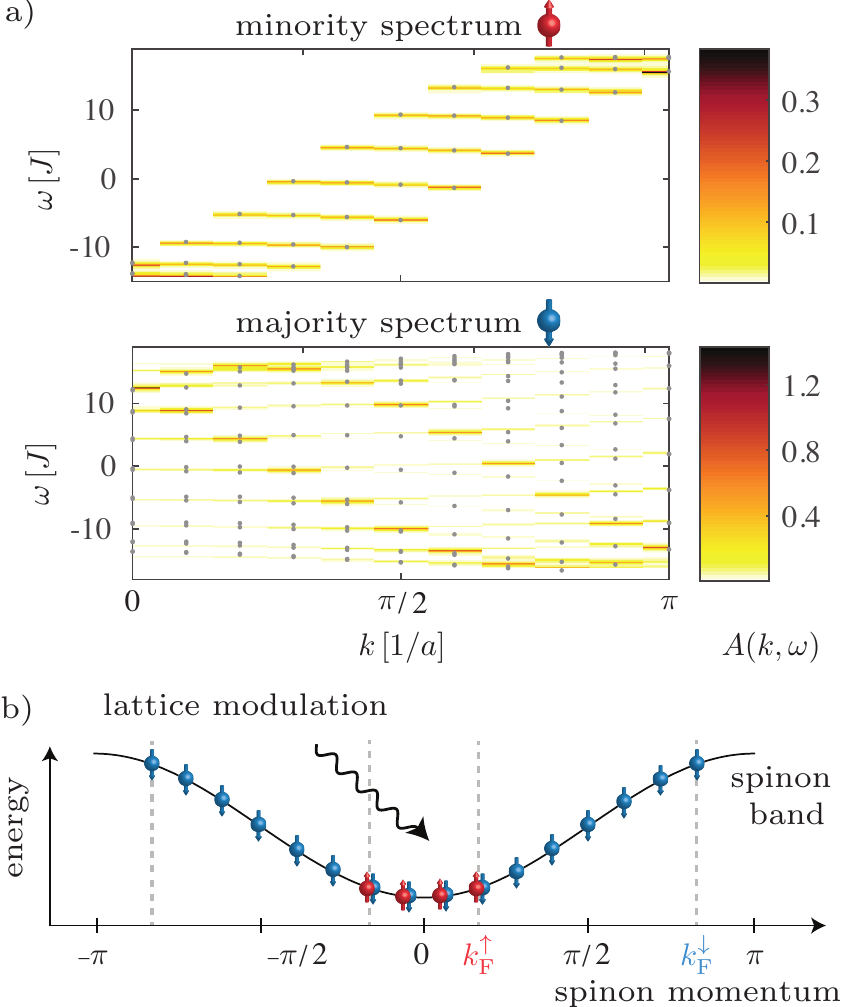, width=0.45\textwidth}
\caption{\textbf{Spectral function in a spin-imbalanced system} with 20 sites and $N_\uparrow = 4$, $N_\downarrow = 16$ at zero temperature and with periodic boundary conditions. a) shows the minority (top) and majority (bottom) spectrum, resolved after the spin of the removed particle. b) depicts the spinon Fermi seas for the two different species, which are filled correspondingly.}
\label{figMagnetized}
\end{figure}

In Fig.~\ref{figMagnetized} a), the spectral function of a single hole in a spin imbalanced system is shown for a removed particle with spin up and down, respectively. As in Fig.~\ref{fig3Temperature}, gray dots denote the positions of expected peaks due to holon and spinon dispersions, Eq.~\eqref{eq:dispersion} for $-k_F^{\uparrow/\downarrow} \leq k \leq k_F^{\uparrow/\downarrow} $ with $k_F^{\uparrow/\downarrow}=\pi N_{\uparrow/\downarrow}/L$. In the slave-fermion mean field theory, the spinons form two Fermi seas, which are filled corresponding to the spin imbalance in the system. Accordingly, in Fig.~\ref{figMagnetized} a), the sharp decrease in spectral weight occurs at different momenta for the removed particle belonging to the majority or minority species.

\section{Theoretical Analysis}
\label{secTheory}
In the previous section we have explained the numerical results for the single-hole spectral function using the semi-phenomenological spectral building principle. We now present a theoretical formalism to obtain the results in Eqs.~\eqref{eq:spinon_dispersion}, \eqref{eq:holon_dispersion} directly from the microscopic Hamiltonian. We describe a single hole in an anti-ferromagnetic spin chain. In order to describe the spin chain, we use a slave-fermion mean field theory \cite{Baskaran1987,Weng1995,Wen2004}, which contains a non-trivial order parameter, that is finite even in one dimension. For simplicity we consider situations with zero total magnetization.

Our starting point is the $t$-XXZ model Eq.~\eqref{eq:tjzmodel} with zero or one hole. We introduce slave boson operators $\h_j$ to describe the holons, and constrained fermions $\f_{j,\alpha}$ describing the remaining spins \cite{Auerbach1998}. The index $\alpha=\uparrow,\downarrow$ corresponds to the two spin states and it holds
\begin{equation}
\hat{\mathbf{S}}_i = \frac{1}{2} \sum_{\alpha, \beta} \f_{i,\alpha}^\dagger \boldsymbol{\sigma}_{\alpha,\beta} \f_{i,\beta}.
\label{eq:spinon_operators}
\end{equation}
The slave particles satisfy the condition
\begin{equation}
\sum_\alpha \fd_{j,\alpha} \f_{j,\alpha} + \hd_j \h_j = 1
\end{equation}
and the original fermionic operators can be expressed as
\begin{equation}
\c_{j,\alpha} = \hd_j \f_{j,\alpha}.
\label{eqCjalpha}
\end{equation}
Using the new operators, one can identify a spin state $\ket{\sigma_1,...,\sigma_L}$ with $\sigma_j = \uparrow, \downarrow$ as
\begin{equation}
\ket{\sigma_1,...,\sigma_L} \equiv  \fd_{1,\sigma_1} ... \fd_{L,\sigma_L} \ket{0}
 \label{eqSigmaDefF}
\end{equation}
and create all states with holes by applying $\hd_j \f_{j,\alpha}$ from Eq.~\eqref{eqCjalpha}. Note that the ordering of $\f$ operators in Eq.~\eqref{eqSigmaDefF} is important due to their fermionic anti-commutation relations. 

We can substantially simplify the formalism by introducing a new basis where the holons occupy bonds between the lattice sites $\tilde{j}$ of the so-called squeezed space \cite{Zaanen2001,Kruis2004a,Hilker2017}, which is obtained by removing all holes from the spin chain. By including only operators $\f_{j,\alpha}$ on sites $\tilde{j}$ we obtain new operators $\tf_{j,\alpha}=\f_{\tilde{j},\alpha}$ with $\tilde{j}= j + \sum_{i\leq j} \hd_i \h_i$. The main advantage of this mapping is the form of the hopping term in Eq.~\eqref{eq:tjzmodel}: using the original operators $\f_{j,\alpha}$, we obtain a difficult quartic expression $- t \sum_\ij \sum_\sigma \hd_j \h_i \fd_{i,\sigma} \f_{j,\sigma}$. In contrast, in terms of the operators $\tf_{j,\alpha}$, a quadratic term involving only holons is obtained, $- t \sum_\ij \hd_j \h_i$, see Appendix \ref{AppdxSBPderivation}. This leaves the spin order in squeezed space unchanged. Moreover, this term yields the dominant part of the free holon dispersion relation Eq.~\eqref{eq:holon_dispersion}, $- 2 t \cos (k_h)$. Corrections due to next-nearest neighbor tunneling, which is included in the $t-J^*$ model, are derived in Appendix \ref{AppdxSBPderivation}.

By creating a hole and removing a spin the number of $\tf$ fermions changes according to Eq.~\eqref{eqCjalpha}. The total spin is thus changed by $1/2$ and the operators $\tf$ can be identified with fermionic spinons \cite{Wen2004}. In squeezed space, the last two terms of Eq.~\eqref{eq:tjzmodel} do not change and correspond to a spin chain without doping. In subsection \ref{secMF} we derive the shape of the spinon dispersion Eq.~\eqref{eq:spinon_dispersion} by considering the undoped spin-chain.

In addition, there exist interactions between the holon and the surrounding spins in squeezed space. As discussed in detail in Appendix \ref{AppdxSBPderivation}, the presence of a holon on the bond between sites $\tilde{j}$ and $\tilde{j}+1$ effectively switches off the coupling between the corresponding spins in the $t$-XXZ model \eqref{eq:tjzmodel}. In the $t-J^*$ model \eqref{eq:tjmodel}, it also affects next-nearest neighbor couplings. We discuss in subsection \ref{subsecHolonPolaron} how these interactions renormalize the holon properties.

\subsection{Slave-fermion mean field theory of \\ undoped spin chains}
\label{secMF}
In this section, we present a slave-fermion mean field theory for the undoped XXZ spin chain which -- up to a pre-factor -- allows us to derive the exact spinon dispersion relation known from Bethe ansatz. Furthermore, it enables an intuitive understanding of the asymmetry in the distribution of spectral weight around $k=\pi/2$ found in the case of isotropic Heisenberg couplings, see Sec.~\ref{subsecSpectralSignaturesLuttSpinLiq}. 

We consider the slave fermion operators $\tf_{j,\alpha}$ discussed above by introducing the notion of squeezed space. 
The Hamiltonian of the spin chain, Eq. \eqref{eq:tjzmodel} at half filling, can be expressed in terms of the spinon operators \cite{Wen2004},
\begin{align}
\begin{split}
\H_{\rm XXZ} = 
-\frac{1}{2} \sum_{i,\alpha} \tf_{i,\alpha}^\dag \tf_{i+1,\alpha}  \left[ J_\perp  \tf_{i+1,\bar{\alpha}}^\dag \tf_{i,\bar{\alpha}}+
 J_z   \tf_{i+1,\alpha}^\dag \tf_{i,\alpha}  \right] 
\\
+ \frac{J_z}{2}\sum_{i,\alpha} \tf_{i,\alpha}^\dag \tf_{i,\alpha} -\frac{J_z}{4}\sum_{i,\alpha,\beta} \tf_{i,\alpha}^\dag \tf_{i,\alpha} \tf_{i+1,\beta}^\dag \tf_{i+1,\beta},
 \label{eqHfsOriginal}
\end{split}
\end{align}
where, $\bar{\uparrow}={\downarrow}$ and $\bar{\downarrow}={\uparrow}$. This expression is exact within the subspace defined by the constraint $\sum_\alpha \tf_{i,\alpha}^\dag \tf_{i,\alpha} = 1$. In the mean field approximation applied below, this constraint is replaced by its ground state expectation value
\begin{equation}
\sum_\alpha \left<\tf_{i,\alpha}^\dag \tf_{i,\alpha}\right> = 1.
\end{equation}

\subsubsection{Mean-field description}
In the following we consider the case of zero total magnetization in the thermodynamic limit. 
At the isotropic point, $J_z = J_\perp$, Eq.~\eqref{eqHfsOriginal} becomes the $SU(2)$ invariant Heisenberg Hamiltonian $ \H_{\rm H}$. In this case we replace the operator $\tf_{i,\alpha}^\dag \tf_{i+1,\alpha}$ by its ground state expectation value 
\begin{equation}
\chi_{i,\alpha} = \left<\tf_{i,\alpha}^\dag \tf_{i+1,\alpha} \right>.
\end{equation}
When $\chi_{i,\alpha} = \chi$ is independent of the spin index $\alpha$, the resulting mean-field Hamiltonian is also $SU(2)$ invariant. By diagonalizing the latter we obtain a self-consistency equation for $\chi$ which will be solved numerically below. 

To obtain a mean-field description away from the $SU(2)$ invariant Heisenberg point, i.e. when $J_z \neq J_\perp$, we can write the original Hamiltonian as a sum of the Heisenberg term $\H_{\rm H}$ plus additional Ising couplings,
\begin{equation}
\H_{\rm XXZ} = \H_{\rm H} +\underbrace{(J_z - J_\perp)}_{= \Delta J_z} \frac{1}{4} \sum_i \hat{\delta}_i \hat{\delta}_{i+1},
\label{eqXXZperturbation}
\end{equation}
where $\hat{\delta}_i =2 \hat{S}^z_i$ is the local magnetization,
\begin{equation}
\hat{\delta}_i = \sum_\alpha (-1)^\alpha  \tf_{i,\alpha}^\dag \tf_{i,\alpha}, \qquad (-1)^\uparrow=1, (-1)^\downarrow=-1.
\end{equation}

We also allow for a finite expectation value of the magnetization in the mean-field description. Assuming that the discrete symmetry $\hat{T}^x \hat{S}^x$, which flips the spins and translates the system by one lattice site, is unbroken, we obtain
\begin{equation}
\langle \hat{\delta}_i \rangle = (-1)^i \delta.
\label{eqMFconstr}
\end{equation}
This leads to a second self-consistency equation for the staggered magnetization $\delta$. 

\begin{figure}[b!]
\centering
\epsfig{file=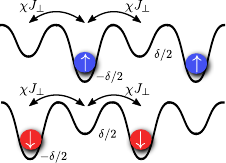, width=0.27\textwidth}
\caption{\textbf{Slave-fermion description of the anisotropic spin chain.} In the anisotropic ${\rm XXZ}$ spin chain, the sub-lattice symmetry can be spontaneously broken when $J_z > J_\perp$. In this case the effective hopping Hamiltonian of spinons corresponds to a tight-binding model with alternating on-site potentials. The mean field solutions for different spins are related by a translation of one lattice site. }
\label{figxMFTising}
\end{figure}

The effective mean-field Hamiltonian is obtained from Eq.~\eqref{eqXXZperturbation} by introducing the order parameters $\delta$ and $\chi$ and keeping terms up to quadratic order. It has a two-site unit cell because the magnetization is opposite for different sub-lattices. This corresponds to a tight-binding Hamiltonian with nearest-neighbor tunneling of strength $J_\perp \chi$ and on-site potentials $(-1)^{i} (-1)^{\alpha} \delta / 2$, as illustrated in Fig.~\ref{figxMFTising}. For spinons of type $\alpha$ it can be written as
\begin{equation}
\H_\alpha =  \int_{-\frac{\pi}{2}}^{\frac{\pi}{2}} dk \left( \tf_{k,A,\alpha}^\dag \tf_{k,B,\alpha}^\dag \right) \underline{\underline{h_\alpha}}(k) \begin{pmatrix}
\tf_{k,A,\alpha} \\ \tf_{k,B,\alpha} \end{pmatrix},
\label{eq:H1}
\end{equation}
where we defined the Fourier transformed spinon operators $\tf_{k}$ by the relations
\begin{align}
\begin{split}
\tf_{2n,\alpha} &=\sqrt{\frac{L}{2\pi}} \int_{-\frac{\pi}{2}}^\frac{\pi}{2} dk ~ e^{-2ikn} \tf_{k,A,\alpha}
\\
\tf_{2n+1,\alpha} &=\sqrt{\frac{L}{2\pi}} \int_{-\frac{\pi}{2}}^\frac{\pi}{2}dk~ e^{-2ikn} e^{-ik} \tf_{k,B,\alpha}. 
\end{split}
\end{align}
For $\alpha = \uparrow$ spinons it holds
\begin{equation}
\underline{\underline{h_\uparrow}}(k) = \begin{pmatrix}
- \nicefrac{\delta}{2} \Delta J_z & -2 \chi J_\perp \cos(k_x) \\ -2 \chi J_\perp \cos(k_x) &  \nicefrac{\delta}{2} \Delta J_z \end{pmatrix},
\label{eqheffectiveMF}
\end{equation}
and a similar expression is obtained for $\alpha = \downarrow$ by changing $\delta \to - \delta$. In addition, there is a constant energy contribution of $J_\perp (\nicefrac{1}{4} + 2 \chi) + \nicefrac{1}{4} \Delta J_z \delta^2$ per particle which is not included in Eq.~\eqref{eq:H1}.

\subsubsection{Self-consistency equation}
To derive the coupled self-consistency equations for $\delta$ and $\chi$, we start by diagonalizing the mean-field Hamiltonian. A new set of spinon operators $\hat{F}_{k,\mu,\alpha}$, with band index $\mu=\pm$, can be defined, for which
\begin{equation}
\H_\alpha = \sum_{\mu = \pm 1} \mu \int_{-\frac{\pi}{2}}^{\frac{\pi}{2}} dk~   \epsilon_k ~  \hat{F}_{k,\mu,\alpha}^\dag \hat{F}_{k,\mu,\alpha}.
\end{equation}
The mean-field dispersion relation is given by
\begin{equation}
\epsilon_k = \sqrt{(2 \chi J_\perp \cos(k))^2 + (\Delta J_z \delta / 2 )^2},
\end{equation}
which gives rise to a band-gap to collective excitations of
\begin{equation}
\Delta_{\rm MF} = |\Delta J_z \delta|.
\label{eqGapSpinons}
\end{equation}
Thus a non-vanishing staggered magnetization $\Delta J_z \delta \neq 0$ opens a gap in the spectrum. Because of the mean-field constraint in Eq.~\eqref{eqMFconstr} we obtain $ \left<\tf_{i,\alpha}^\dag \tf_{i,\alpha}\right> = 1/2$, i.e. we describe spinons at half filling. When $\Delta J_z \delta \neq 0$ the ground state is a band insulator, whereas $\Delta J_z \delta = 0$ corresponds to a gapless spinon Fermi sea. 

Using the new spinon operators $\hat{F}_{k,\mu,\alpha}$ we can calculate the order parameters $\delta$ and $\chi$ self-consistently,
\begin{flalign}
\chi &= \frac{1}{\pi} \int_{-\frac{\pi}{2}}^\frac{\pi}{2} dk ~ \cos^2(k) \frac{\chi J_\perp}{\epsilon_k},
\label{eq:chiA1_a}
\\
\delta &=  \frac{1}{2 \pi} \int_{-\frac{\pi}{2}}^\frac{\pi}{2} dk ~ \frac{\Delta J_z \delta }{\epsilon_k}.
\label{eq:chiA1_b}
\end{flalign}

\begin{figure}[t!]
\centering
\epsfig{file=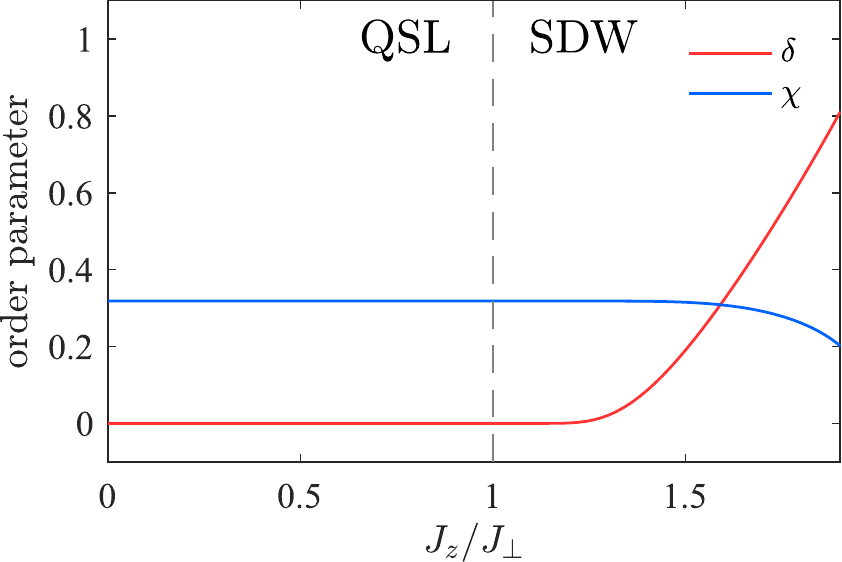, width=0.43\textwidth}
\caption{\textbf{Mean field theory for the spin chain.} Numerical solution of the self-consistency equations for the order parameters $\chi$ and $\delta$, Eqs.~\eqref{eq:chiA1_a}, \eqref{eq:chiA1_b}. For $0 \leq J_z \leq J_\perp$ the ground state is a gapless quantum spin liquid (QSL). For $J_z \gtrsim J_\perp$ the two order parameters $\delta$ and $\chi$ are both non-vanishing and the ground state is a spin-density wave (SDW).}
\label{figxMFT}
\end{figure}

\subsubsection{Mean field phase diagram and singular phase transition}

The numerical solutions for $\delta$ and $\chi$ to Eqs.~\eqref{eq:chiA1_a}, \eqref{eq:chiA1_b} are shown in Fig.~\ref{figxMFT}. For $0 \leq J_z \leq J_\perp$, the only solution is the symmetric one with $\delta=0$ and $\chi=1/\pi$ and energy $E_{\rm MF}(\delta=0,\chi=1/\pi)$. This state is paramagnetic and corresponds to a quantum spin liquid (QSL). At the isotropic Heisenberg point where $J_z=J_\perp = J$ it predicts the following dispersion of spinon excitations,
\begin{equation}
\epsilon_k = J \frac{2}{\pi} |\cos k|.
\end{equation}
The analytical form of the spinon dispersion $\sim |\cos (k)|$ is correctly described by the mean-field theory. Compared to the exact result from Bethe ansatz calculations, Eq.~\eqref{eq:spinon_dispersion}, this expression is too small by a factor of $\pi^2/4 \approx 2.47$. Deviations from the exact solution are a result of the mean field approximation, i.e. our neglecting of gauge fluctuations ensuring the constraint of single-occupancy \cite{Wen2004}.

For $J_z > J_\perp$, two additional solutions $\pm \delta \neq 0$ with an energy below $E_{\rm MF}(\delta=0,\chi=1/\pi)$ appear (only the solution with $\delta >0$ is shown in Fig.~\ref{figxMFT}). In this regime, the translational symmetry of the original Hamiltonian Eq.~\eqref{eqHfsOriginal} is spontaneously broken. Because there exists a non-zero staggered magnetization $\delta \neq 0$, this phase can be identified with a spin density wave (SDW). At large couplings, $J_z \geq 2 J_\perp$, we find that the mean-field order parameter $\chi$ vanishes and the system is fully ordered with $\delta = \pm 1$ as expected in the classical N\'eel state. This second transition is an artifact of the mean-field theory: From exact Bethe ansatz calculations it is known that the staggered magnetization approaches the classical value $\delta = \pm 1$ monotonically until it is asymptotically reached for $J_z/J_\perp \to \infty$. Here, we are more interested in the behavior of the transition at $J_z = J_\perp$. 

As can be seen from Fig.~\ref{figxMFT}, the order parameter $\delta$ only takes a significant value for $J_z \simeq 1.2 J_\perp$. By solving the elliptic integral in Eq.~\eqref{eq:chiA1_b} perturbatively in the limit $\delta \ll 1$, we find that the staggered magnetization depends non-analytically on $J_z - J_\perp$, with all derivatives $d^n \delta/ d \Delta J_z^n = 0$ vanishing at the Heisenberg point:
\begin{equation}
\delta \simeq \frac{4}{\pi} \frac{J_\perp}{J_z - J_\perp} e^{-2 \frac{J_\perp}{J_z - J_\perp}}.
\end{equation}
From Eq.~\eqref{eqGapSpinons} it follows that the excitation gap has the asymptotic form
\begin{equation}
\Delta_{\rm MF} \simeq \frac{4}{\pi} J_\perp \exp \left[ - 2 \frac{J_\perp}{J_z - J_\perp} \right].
\end{equation}

The excitation gap $\Delta_{\rm MF}$ close to the transition point from QSL to SDW can be compared to exact results $\Delta_{\rm B}$ obtained from Bethe ansatz methods for the XXZ chain. From the exact expressions derived in Ref.~\cite{Sutherland} we obtain the following asymptotic behavior,
\begin{equation}
\Delta_{\rm B} \simeq 4 \pi J_\perp \exp \left[ - \frac{\pi^2}{2\sqrt{2}} \l \frac{J_\perp}{J_z - J_\perp} \r^{1/2} \right].
\end{equation}
The non-analyticity is correctly predicted by the mean-field theory, and only the power-law exponent appearing in the exponential function is not captured correctly. 

We conclude that the slave-fermion mean field theory provides a rather accurate description of the one-dimensional spin chain near the critical Heisenberg point. This is possible because a non-trivial order parameter ($\chi$) is introduced that does not vanish even in one dimension. The theory provides quantitatively reasonable results and describes correctly the qualitative behavior at the singular phase transition from QSL to the conventional symmetry broken SDW phase. We now show that it moreover offers a simple explanation of the observed asymmetric spectral weight in the single-hole spectral function of the Heisenberg spin chain.

\subsubsection{Spectral weight of spinon excitations}

We proceed by calculating the matrix elements that determine the weight in the single-hole spectra based on the slave-fermion mean field theory. The relevant matrix elements are of the form
\begin{equation}
\lambda_{k_s}^n = | \bra{\psi_n} \tf_{k_s,\sigma} \ket{\psi_0}|^2 \qquad -\pi \leq k_s \leq \pi,
\end{equation}
that describe the creation of a hole in the ground state of the spinon system. Here,
\begin{equation}
\ket{\psi_0} = \prod_{k=-\nicefrac{\pi}{2}}^{\nicefrac{\pi}{2}} \prod_\sigma \hat{F}_{k,-,\sigma}^\dag \ket{0}
\end{equation}
is the ground state of the undoped spin chain.

The full spectral function $A(k,\omega)$ is a convolution of the spinon part and the holon part,
\begin{multline}
A(k,\omega) = \sum_{k_h,k_s=-\pi}^\pi \int d\omega_h d\omega_s ~ \delta\l \omega - \omega_{\rm s} - \omega_{\rm h} \r \\
\times  \delta_{k,k_h+k_s}  A_s\l k_s , \omega_{\rm s} \r A_h(k_h,\omega_h).
\end{multline}
Neglecting the coupling of the holon to collective excitations of the spin chain, see Sec.~\ref{subsecHolonPolaron}, the holon spectrum is determined by $A_h(k_h,\omega_h) = \delta(\omega_h-\epsilon_h(k_h))$. The spinon part is given by $A_s(k_s,\omega_s) = \sum_n \delta(\omega_s - \omega_n) \lambda_{k_s}^n$, where the eigenstate $\ket{\psi_n}$ has energy $\omega_n$. 

For every $k_s$ there exists one unique state $\ket{\psi_n}$ with $\lambda_{k_s}^n \neq 0$. The corresponding $\lambda_{k_s} := \lambda_{k_s}^n$ can be calculated by mapping the original spinon operators $\tf_{k_s,\sigma}$ onto the transformed operators $\hat{F}_{k_s,\pm,\sigma}$. This leads to
\begin{align}
\lambda_{k_s} = \begin{cases} \cos^2 \l \frac{\theta_{k_s}}{2} \r	\qquad |k_s|\leq \pi/2
\\
\sin^2 \l \frac{\theta_{k_s}}{2} \r	\qquad |k_s| > \pi/2,
\end{cases}
\end{align}
where the mixing angle is determined by
\begin{equation}
\tan \theta_{k_s} = \frac{\delta (J_z - J_\perp)}{4 \chi J_\perp \cos(k_s)}
\end{equation}

In the isotropic Heisenberg case, $J_\perp = J_z = J$, the only solution to the self-conistency equation is $\delta = 0$, leading to $\theta_{k_s} = 0$ and thus 
\begin{align}
\lambda_{k_s} = \begin{cases} 1	\qquad |k_{s}|\leq \pi/2
\\
0	\qquad |k_s| >\pi/2.
\end{cases}
\end{align}
This discontinuity in $\lambda_{k_s}$ gives rise to the sharp drop of spectral weight observed in Fig.~\ref{fig1Setup} b), Fig.~\ref{fig3Temperature} and Fig.~\ref{figIsing} a) when $k_s$ is varied across the value $\pi/2$. It is a direct signature for the spinon Fermi sea, which in turn is a key signature of a quantum spin liquid. 

In the Ising limit $J_\perp = 0$ we obtain the classical N\'eel state with $\delta = \pm 1$ and $\chi=0$. This yields $\theta_{k_s}=\pi/2$, i.e. 
\begin{equation}
\lambda_{k_s} = \nicefrac{1}{2}.
\end{equation}
In this case, discrete translational symmetry is broken, which leads to a mixing of momenta $k_s$ and $k_s + \pi$ and a homogeneous re-distribution of spectral weight across all $k_s$. There is therefore no discontinuity in the distribution spectral weight at the zone boundary $k_s = \pm \pi/2$.

\subsection{Renormalization of holon properties: \\ the holon-polaron}
\label{subsecHolonPolaron}

In our analysis of the single-hole spectrum in Sec.~\ref{subsecSpinChargeSep}, based on the spectral building principle, we neglected couplings of the holon to the spin environment. We now discuss leading order corrections to this picture, which scale as $J/t$.~Experimentally, the relevant parameter regime of the $t-J^*$ model is $J/t \ll 1$, therefore these corrections are generically small.

The essence of spin-charge separation is that the spinon and the holon are not bound to one another and can be treated independently. Nevertheless, when the holon is moving through the spin chain, it interacts with the surrounding spins and becomes dressed by collective excitations with vanishing total spin. This effect can be understood by the formation of a polaronic quasiparticle \cite{Landau1948}, which we will refer to as the holon-polaron from now on. Note that this situation is different from two dimensions. In that case there is no spin-charge separation and a magnetic polaron carrying spin $1/2$ is formed by a hole moving in a two-dimensional N\'{e}el state \cite{Bulaevskii1968,Brinkman1970,SchmittRink1988,Kane1989,Sachdev1989,Liu1991,Mishchenko2001,Manousakis2007}. 

We start from the $t-J^*$ Hamiltonian \eqref{eq:tjmodel} which is an exact asymptotic representation of the Fermi-Hubbard model for large $U$ at half filling. Then we use a formulation in squeezed space \cite{Kruis2004a}, where the holon effectively moves between the bonds of the lattice on which it switches off the superexchange interaction. The collective excitations of the spin chain, which can be understood as particle-hole pairs $\tfd \tf$ in the spinon Fermi sea discussed in the previous section, are then described using the bosonization formalism \cite{Haldane1981} and assuming an infinite system. In combination, we arrive at a conventional polaron Hamiltonian that can be solved perturbatively for weak polaronic couplings $J \ll t$. Note that weak polaronic coupling corresponds to large coupling $U \gg t$ in the original Fermi Hubbard model. For details of our calculations we refer to Appendix \ref{secPolaronDescription}.

We calculated the leading-order corrections to the holon-polaron properties. For the ground state energy of the holon-polaron we obtain
\begin{equation}
E_{\rm h}^0 = - 2 t - \frac{J}{4} - \frac{J^2}{t} \left[ 0.0343 + 6.54 m^2 + 5.31 |C|^4 \right]. \label{eq:EhPiComp}
\end{equation}
This energy is measured relative to a chain with the same number of spins but without the holon. Here $m=(N_\uparrow-N_\downarrow)/2L$ denotes the magnetization per length. The non-universal constant $|C|^2\approx 0.14$ was determined by Eggert and Affleck \cite{Eggert1995} from comparison of the spin-structure factor obtained from bosonization and quantum Monte Carlo calculations. 

The effective mass of the holon polaron is defined by expanding its energy $E_{\rm h}(p_{\rm h})$ around momentum $p_{\rm h}=0$ where $E_{\rm h}$ is minimized,
\begin{equation}
E_{\rm h}(p_{\rm h}) = E^0_{\rm h} + \frac{1}{2 M_{\rm h}} p_{\rm h}^2 + \mathcal{O} p_{\rm h}^4.
\end{equation}
For the renormalized holon mass we obtain
\begin{multline}
\frac{1}{M_{\rm h}} = 2 t - 2.77 J + 39.5 J m^2 \\
- \frac{J^2}{t} \left[ 0.188 + 87.2 m^2 + 43.2 |C|^4 \right]. \label{eq:MhPiComp}
\end{multline}
The expressions \eqref{eq:EhPiComp} and \eqref{eq:MhPiComp} are correct up to terms of order $\mathcal{O}(J^3/t^2)$.

For parameters as in Fig.~\ref{figIsing} a), i.e. $t=8 J$ and $m=0$, we obtain corrections to the holon energy of $\Delta E_{\rm h}^0 = E_{\rm h}^0 + 2 t = - 0.27 J$. The ground state energy per bond in the spin-chain without the hole is $E_0 / L \approx  - 0.44 J$, see e.g. Ref.~\cite{Giamarchi2003}. Hence we expect the lower edge of the spectrum at $\omega_- = - 2 t + (0.44 - 0.27) J = -15.83 J$. The corrections are of the correct order of magnitude, as can be seen by comparison to the value $\omega_- \approx -15.80J$, which has been obtained from a finite size scaling of exact diagonalization results~\cite{fss}. We expect that the dominant source for errors are finite size effects and the ambiguity of the ultra-violet cut-off chosen in the bosonization. In the context of Bose polarons in one dimension it has been shown that the latter effect can lead to sizable corrections to bosonization results \cite{Grusdt2016RG1D}. For the renormalized mass we obtain $2 t M_{\rm h} = 1.22$, which corresponds to a $22 \%$ mass enhancement. This value is consistent with the exact numerical results in Fig.~\ref{figIsing}, but it is too small for a meaningful direct comparison due to finite-size effects and in particular the required momentum resolution.

In principle both the holon mass and energy renormalization can be measured experimentally by close inspection of the spectrum. However, as demonstrated above, the overall effect is very weak. Using ultracold atoms, the $t-J$ model can also be implemented independently of the Fermi-Hubbard model by using polar molecules \cite{Gorshkov2011tJ} or Rydberg dressing \cite{Zeiher2016,Zeiher2017}. This allows one in principle to tune the polaronic coupling $J/t$ to arbitrary values, smaller or larger than one. When $J \gg t$, we expect a strong renormalization of the holon-polaron properties which can be studied in the future using a formalism along the lines of the one presented in Appendix \ref{secPolaronDescription}. 

In contrast to the situation for the $t-J^*$ model, the limit $J/t \to 0$ is not well defined for the $t-J$ model. We show in Appendix \ref{secPolaronDescription} that one obtains infrared-divergent integrals to lowest order for the energy and the effective mass because the next-nearest neighbor terms are missing.

\section{Extensions}
\label{secExtensions}

The scheme for measuring the spectral function of a single hole can be generalized to implement different spectroscopic probes using ultracold atoms. In this section we briefly illustrate two examples, although a detailed analysis is devoted to future work. We show how the dynamical structure factor $S(\omega,k)$ can be measured in one-dimensional spin chains (\ref{subsecSkw}), and discuss how the scheme can be extended to implement the analog of double photoelectron spectroscopy \cite{Herrmann1998,Fominykh2000} (\ref{subsecDPE}).

\subsection{Dynamical spin structure factor}
\label{subsecSkw}

\begin{figure}[t!]
\centering
\epsfig{file=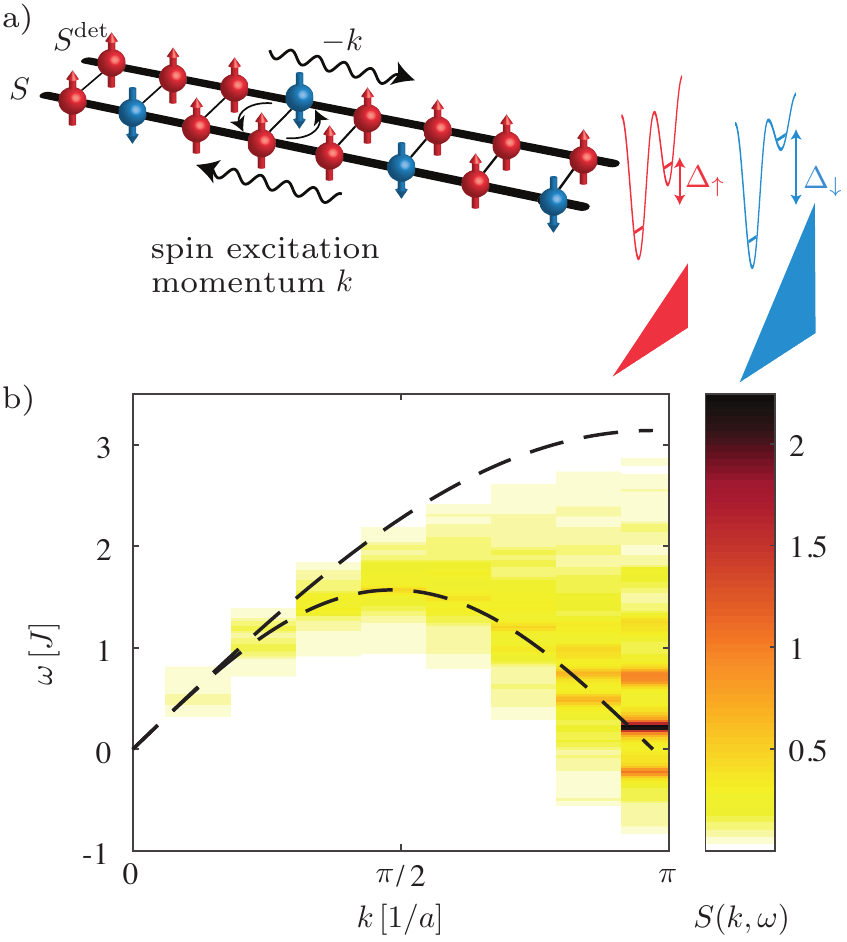, width=0.43\textwidth}
\caption{\textbf{Direct measurement of the dynamical spin structure factor.} a) The empty probe system is replaced by a fully polarized spin chain and coupled to the system using superexchange interactions in the $y$-direction. By detecting a single magnon created in $S^{\rm det}$ and measuring its momentum, the dynamical spin structure factor $S(k,\omega)$ can be measured. In b) we calculate $S(k,\omega)$ for the isotropic Heisenberg model in a finite-size box of length $L=16$ at a temperature $T=0.2J$. It shows the expected broad continuum reflecting the fractionalization of a spin-flip excitation into a pair of spinons. The black dashed lines correspond to the theoretically expected upper and lower boundaries at zero temperature.}
\label{figSkw}
\end{figure}

The spectral function $A(k,\omega)$ probes the properties of a single hole interacting with the surrounding spins. To obtain information about the spin system alone, more direct measurement schemes are required where no charge excitations are generated. The most common example is the dynamical spin structure factor $S(k,\omega)$, where a spin-flip excitation with momentum $k$ is created at an energy $\omega$. Using a Lehmann representation similar to Eq.~\eqref{eq:spectralfunction} it can be defined by
\begin{multline}
S(k,\omega)= \frac{1}{2 Z_0} \sum_{n,m}  e^{-\beta E_n^{M}} |\bra{\psi_m^{M+1}} \hat{S}^+_{k} \ket{\psi_n^{M}} |^2 \\
\times \delta (\hbar \omega - E_m^{M+1} + E_n^{M}),
\label{eqLehmanSkw}
\end{multline}
with $\ket{\psi_n^M}$, $E_n^M$ denoting the eigenstates and -energies of the system $S$ with total magnetization $M$. In solids, $S(k,\omega)$ can be measured in inelastic neutron scattering experiments~\cite{Nagler1991}. 

To directly measure the dynamical spin structure factor $S(k,\omega)$ using a quantum gas microscope, we propose to replace the empty probe system $S^{\rm det}$ by a spin chain which is fully polarized along the $z$-direction, see Fig.~\ref{figSkw} a). Instead of the modulated tunnel coupling $\hat{T}_y$ from Eq.~\eqref{eq:pert}, the system can be coupled to the probe by super-exchange interactions,
\begin{equation}
\hat{J}_y = \sum_i \hat{\vec{S}}_{i} \cdot  \hat{\vec{S}}_{i, \rm det},
\end{equation}
where $\hat{\vec{S}}_{i, \rm det}$ denotes the spin operator on site $i$ in the probe system. By modulating the tunneling amplitude $t_y$ as in the case of the spectral function, the resulting super-exchange coupling $j_y$ is also modulated with the same frequency $\omega_{\rm shake}$. The time-dependent perturbation thus reads $\H_{\rm pert}(\tau) = \delta j_y \sin (\omega_{\rm shake} \tau) \hat{J}_y$, cf. Eq.~\eqref{eq:Hpert}.

Similar to the case of the spectral function, the perturbation creates an excitation in the probe system $S^{\rm det}$: here, the excitation is a magnon carrying spin $S^z=-1$ with momentum $-k$ and energy $\omega_m(-k)$, where $\omega_m(k)$ denotes the magnon dispersion relation. In order to measure the momentum of the magnon, we assume that spin up and down states have different magnetic moments. By applying a magnetic field gradient along the $x$-direction before taking a spin-resolved image \cite{Boll2016} one can thus implement the Wannier-Stark mapping discussed in Sec.~\ref{subSecMomentum} for the single magnon in $S^{\rm det}$. Similarly, a magnetic field gradient along the $y$-direction can be used to realize spin-dependent energy offsets $\Delta_\sigma$, analogous to the energy offset $\Delta$ considered in the case of the spectral function. Finally, by measuring the position of the magnon after applying the Wannier-Stark mapping, the excitation rate $\Gamma(k, \omega)$ is obtained, which is directly related to the dynamical spin structure factor
\begin{equation}
 \Gamma(k, \omega)= \frac{2\pi}{\hbar} |\delta j_y|^2 S(k, \omega),
\end{equation}
as obtained by Fermi's golden rule. 

In Fig.~\ref{figSkw} b) we show an example for the dynamical spin structure factor $S(k,\omega)$, calculated at finite temperature and for realistic system sizes accessible in current experiments. The spin-flip creates a pair of two fractionalized spinon excitations. As a result, one can observe a broad spinon continuum, which is considered a key indicator for a quantum spin liquid.

\subsection{Double photoelectron spectroscopy}
\label{subsecDPE}
Further insight into the nature of excitations in the system can be obtained by measuring their spatial correlations. An interesting method which achieves this goal in solids is double photoelectron spectroscopy \cite{Herrmann1998,Fominykh2000}, where a correlated pair of two photoelectrons is emitted and detected. Similarly, processes can be considered where two atoms are transferred into the initially empty probe system $S^{\rm det}$. To avoid interactions between the two emitted atoms, one could e.g. consider a situation with two probe systems $S^{\rm det}_{\rm L,R}$, one to the left and one to the right of the system $S$, and post-select on cases with one atom per probe system. 

The resulting spectrum contains pairs of individual one-particle events as well as two-particle processes which provide additional information about the system. The two-particle contributions can be distinguished from one-particle effects by using coincidence measurements. In this technique one post-selects events where both excitations are created simultaneously. In the quantum gas microscope setups discussed here this can be achieved by extending $S^{\rm det}_{\rm L,R}$ in the $y$-direction. One can use the travelled distance from the system $S$ in $y$-direction as a measure of the time that passed between the creation and the detection of a particle. 

From the coincidence measurements described above, information about the two-hole spectral function $A_{12}(k_1, k_2, k_1',k_2'; \omega)$ can be extracted, see Ref.~\cite{Fominykh2000}. It contains information about the correlations between the two created holes in the system. These correlations are expected to be weak in a system with one-dimensional chains with spin-charge separation, where holons form a weakly interacting Fermi sea \cite{Ogata1990,Giamarchi2003}. On the other hand, in systems that are superconducting, correlations are expected to play an important role and give rise to distinct features of Cooper pairing in the two-hole spectrum \cite{Kouzakov2003}. Using ultracold atoms, situations with attractive Hubbard interactions $U<0$ have been realized \cite{Mitra2017} which become superconducting at low temperatures. Here the method described above could be applied to directly access the strong two-particle correlations present in this system.

\section{Summary and Outlook}
\label{secOutlook}

In this work we have proposed a measurement scheme for the single-particle excitation spectrum $A(k,\omega)$ in a quantum gas microscope. Our method can be understood as an analog of angle-resolved photoemission spectroscopy (ARPES), which has been key to the study of excitation spectra in many strongly correlated materials. In our method, the weak tunnel coupling from the system under investigation to an initially empty detection system is modulated with frequency $\omega$. The spectral function $A(k,\omega)$ can then be obtained directly from the tunneling rate into a single-particle eigenstate of the detection system with momentum $k$ and energy $\epsilon(k)$. 

We have analyzed the scheme for single-hole spectra in one-dimensional spin chains. Effects from finite size, non-zero temperature, and the presence of sharp edges are included in our numerical simulations.
We discussed two characteristic features of the spectral function $A(k,\omega)$: (i) spin-charge separation and (ii) the distribution of the spectral weight. (i) Spin-charge separation can be identified in the spectral function both for  isotropic and anisotropic spin couplings. It reveals that a hole created in the system separates into a spin-less holon and a spinon. Moreover, their different characteristic energy scales $t$ and $J$, respectively, can be resolved. 
(ii) For the isotropic  Heisenberg model, we observed a strong suppression of the low-energy spectral weight at momenta $k>\pi/2$. The ground state of the isotropic Heisenberg model is a quantum spin liquid with gapless excitations \cite{Giamarchi2003}. We discussed a slave-fermion mean field description of this state, which can be qualitatively understood as a Fermi sea of non-interacting spinons with Fermi momentum $\pi/2$~\cite{Wen2004}. This description can explain the sharp decrease of the spectral weight when the spinon momentum $k_{\rm s}$ crosses the corresponding spinon Fermi momentum at $\pi/2$.

While related ARPES measurements of the spectral function in solids have already shown distinct spinon and holon peaks at low energies \cite{Kim2006}, an observation with ultracold atoms would allow to distinguish spinon and holon dispersions on all energy scales, because  phonon contributions and effects from higher bands are absent. Moreover, the sharp decrease of the spectral weight at $\pi/2$ is present at temperatures currently achievable in experiments with ultracold fermions. Hence, it could provide the first direct signature of a Luttinger spin liquid of cold atoms. 

The sharp step in the spinon spectral weight across its Fermi momentum is to some extent reminiscent of the Fermi arcs observed in the pseudogap phase of quasi-2D cuprates. To explore the relation of these two phenomena experimentally, ultracold atoms can be used to study the dimensional crossover between the 1D and 2D Fermi Hubbard model in the future. In two dimensional systems with long-range anti-ferromagnetic order it is expected that spinon and holon are bound together in a confined phase \cite{Bulaevskii1968,Brinkman1970,SchmittRink1988,Kane1989,Sachdev1989,Liu1991}, similar to mesons, which are bound states of two quarks \cite{Beran1996,Grusdt2017scDyn}. The transition to the pseudogap phase at finite doping and the microscopic origin of Fermi arcs observed in ARPES is poorly understood. Our method for measuring the spectral function can be generalized to two dimensions, where a second layer can be utilized as the probe system. 

Much of the physics discussed in this article can be related to simple models of non-interacting spinon and holon slave particles. This approach is successful due to a large separation of energy scales associated with holon and spinon dynamics ($t$ and $J$ respectively). When $t$ and $J$ become comparable, however, corrections to the simple physical pictures become relevant. We systematically studied leading order corrections in $J/t$ for the Fermi-Hubbard model at strong coupling. A bosonization formalism was used to describe the holon dressing by collective excitations of the spin chain, and we have derived expressions for the renormalized holon energy and its effective mass. 

An interesting future direction of research is the study of the $t-J$ Hamiltonian with a single hole by tuning the ratio $t/J$. The holon can be understood as a mobile impurity, with a tunable bare mass given by $\sim1/2t$, which is interacting with collective excitations of the spin chain. This allows one to explore connections with one-dimensional impurity problems, for which rich physics have been found close to \cite{Casteels2012,Catani2012,Bonart2012,Fukuhara2013,Grusdt2016RG1D,Volosniev2017} and far from equilibrium \cite{Mathy2012,Knap2014,Meinert2017}.

\section*{Acknowledgements}
The authors would like to thank Christie Chiu and Geoffrey Ji for carefully reading our manuscript and for providing useful comments. The authors also acknowledge fruitful discussions with Gregory Astrakharchik, Immanuel Bloch, Sebastian Eggert, Markus Greiner, Christian Gross, Timon Hilker, Randall Hulet, M\'{a}rton Kan\'{a}sz-Nagy, Salvatore Manmana, Efstratios Manousakis, Matthias Punk, Subir Sachdev, Guillaume Salomon, Richard Schmidt, Imke Schneider, Maksym Serbyn, Yulia Shchadilova, Tao Shi, Yao Wang and Johannes Zeiher. We acknowledge support from the Technical University of Munich - Institute for Advanced Study, funded by the German Excellence Initiative and the European Union FP7 under grant agreement 291763  (AB, MK), the DFG grant No. KN 1254/1-1 (AB, MK), the Studienstiftung des deutschen Volkes (AB), the Harvard Quantum Optics Center and the Swiss National Science Foundation (DG), the Gordon and Betty Moore foundation (FG) and from Harvard-MIT CUA, NSF Grant No. DMR-1308435 as well as AFOSR Quantum Simulation MURI, AFOSR grant number FA9550-16-1-0323 (ED).

\appendix

\section{Implementation of the measurement scheme for the $t-J^*$ model}
\label{Appdx1exptjmodel}

A balanced two-component spin mixture of ultracold fermionic atoms in an optical lattice allows for a clean implementation of the $t-J^*$ model introduced in Eq.~\eqref{eq:tjmodel} in the limit of large $U/t \gg 1$. To create the optical lattice configuration necessary for the detection scheme we propose a standard retro-reflecting laser configuration along the $x$-direction with a lattice depth of $V_x$ and tunneling $t$, and a superlattice configuration in the $y$-direction that creates several copies of decoupled double-well systems, see Fig.~\ref{fig1Setup} a). This has the advantage of obtaining several measurements per experimental cycle. However, a standard lattice along the $y$-direction could also be used and the energy offset $\Delta$ could be created with a digital micro-mirror device. 

The superlattice potential can be created for example by two retro-reflected laser beams at wavelengths $\lambda_y/2$ and $\lambda_y$ \cite{Foelling2007}, which create a short and long wavelength lattice of depth $V_{y}^l$ and $V_{y}^s$. By setting their phase difference $\varphi$ close to $\pi/2$ a controlled energy offset between the two sites of the double well can be introduced with bare tunneling $t_y$. The total optical potential is given by
\begin{eqnarray}
V(x,y) & = & V_x \cos^2\left(2\pi x/\lambda_x\right) \label{eq:optical_potential}
\\
& & + V_{y}^l \cos^2\left(2\pi y/\lambda_y\right) +  V_{y}^s \cos^2\left(4\pi y/\lambda_y-\varphi\right).\nonumber
\end{eqnarray}
The lattice depths along the $y$-direction can be chosen sufficiently deep, such that the tunneling between different double wells is negligible. In addition, the energy offset is much larger than all other energy scales $\Delta \gg U, t$ (but smaller than the energy gap to the next band) to make direct tunneling processes off resonant. This also ensures that there are no atoms in $S^{\mathrm{det}}$ when loading the fermionic spin mixture from the initial harmonic trap into the lattice. 

To implement the detection scheme for the spectral function, we propose to periodically modulate the depth of the long wavelength lattice according to $V_{y}^l(t)=V_{y}^l + \delta V_{y}^l \sin(\omega_{\mathrm{shake}} \tau)$. This leads to an induced oscillatory tunnel coupling $\delta t_y$ along the $y$-direction between the spin system $S$ and the detection system $S^{\mathrm{det}}$. Thereby the perturbation described in Eq.~\eqref{eq:pert} can be realized. The strength of the induced tunneling is given by $\delta t_y = \delta V_{\lambda}\int w_{\mathrm{L}}^{*}(y) \cos^2(2\pi y/ \lambda) w_{\mathrm{R}}(y)dy$, where $w_{\mathrm{L}}(y)$ and $w_{\mathrm{R}}(y)$ denote the Wannier functions of the left and right lattice sites of the double-well system created by the lattices along the $y$-direction \cite{Chen2011Shaking}.

\section{Momentum shifts for the spectral building principle}
\label{AppdxShifts}

In this Appendix, we discuss the different shifts for holon and spinon momenta that have to be taken into account in the application of the spectral building principle in Eqs.~\eqref{eq:dispersion} and~\eqref{eq:dispersion_ising}. 

Strictly speaking, there is no translational invariance in squeezed space. However, it is a good approximation up to $1/L$. The first momentum shift we discuss is due to these corrections. The holon moves along $L$ lattice sites. Its momentum is therefore quantized in units of $2\pi/L$. By contrast, there are only $L-1$ spins, such that the spinon momentum is $k_s = n_s 2\pi/(L-1)$ with $n_s$ integer. Therefore, we have to shift the momentum to account for the different quantization conditions for spinon and holon. The spinon momentum $k_s=k-k_h$ in Eq.~\eqref{eq:dispersion} is thus replaced by 
\begin{equation}
k_s'= k-k_h+\frac{k-k_h}{L-1}.
\label{eqksPrime}
\end{equation}
This is the smallest possible shift to obtain the correct quantization of $k_s'$ as an integer multiple of $2\pi/(L-1)$.

In a chain with periodic boundary conditions, a further momentum correction arises: the holon shifts the spins by one lattice site every time it moves across the entire system. When the spins are translated by one site, the wavefunction picks up an overall phase $e^{i P_s}$ where $P_s$ is the total momentum of the spin chain after removing the holon. $P_s = P_0 + k_s$ is determined from the spinon momentum $k_s$ up to an additive constant $P_0$ which is independent of the momentum. Thereby, a \textit{twisted periodic boundary effect} with twist angle given by the spinon momentum $P_s$ is introduced. This corresponds to a shift of all holon momenta $k_h$ by the momentum $P_s$ of the spins divided by the system size $L$. For periodic boundary conditions, we thus have to replace $k_h$ in Eq.~\eqref{eq:dispersion} with 
\begin{equation}
k_h'=k_h- \frac{P_s}{L}.
\label{eqkhPrime}
\end{equation}
In Fig.~\ref{fig3Temperature} a) and b) we included the corrections from Eq.~\eqref{eqksPrime}, \eqref{eqkhPrime} for the positions of the gray dots and found by comparison to our numerical calculations that $P_0=\pi$. 

At small but finite temperatures, there exist collective excitations which carry momentum $\pi$. They contribute to $P_s$ in Eq.~\eqref{eqkhPrime} and thus shift the holon momentum in the case of periodic boundary conditions by an additional amount of $\pi/L$, 
\begin{equation}
k_h'=k_h-\frac{2\pi+k_s}{L}.
\end{equation}
Correspondingly, additional peaks appear between the ones found at zero temperature, which are marked by blue circles in Fig.~\ref{fig3Temperature} a).

\section{The spectral function of the Majumdar-Ghosh model}
\label{Appdx1MG}

\begin{figure}[b!]
\centering
\includegraphics[width=0.5\textwidth]{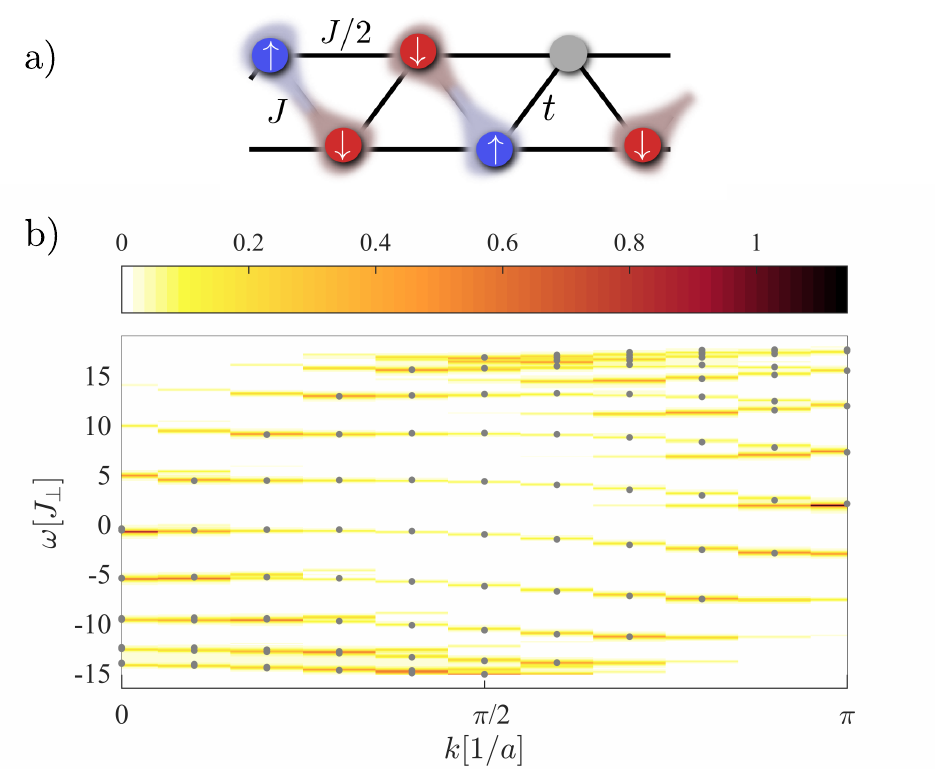}
\caption{\textbf{The Majumdar-Ghosh spin chain.} a) Visualization of the model realized on a zig-zag chain. Next-nearest neighbor hopping terms for the hole are not included here. In b), the spectrum of a Majumdar-Ghosh spin chain with 20 sites and periodic boundary conditions with hopping $t=8J$ and at temperature $T=0$ is shown. Gray dots correspond to spinon and holon dispersion, see Eq. \eqref{eq:spinon_dispersion_MG}.}
\label{figxMG}
\end{figure}

We would like to gain a deeper understanding how the nature of the ground state in a spin chain is related to the single-hole spectral function. Many models underlying frustrated quantum spin systems can be described in terms of resonating valence-bond states \cite{Auerbach1998}. To understand how the valence-bond nature influences the spectral function, we study the Majumdar-Ghosh model. Its ground state can be represented exactly using resonating valence-bond states.

By adding next-nearest neighbor couplings for the spins to Eq.~\eqref{eq:tjzmodel}, we arrive at the Hamiltonian of the Majumdar-Ghosh model \cite{Majumdar1969a,Majumdar1969,Auerbach1998} interacting with a hole-like impurity described by $\h_j$,
\begin{multline}
\H_{\rm MG} = t \sum_j \l \hd_{j+1} \h_j + \hc \r +  J \sum_i \hat{\mathbf{S}}_i \cdot \hat{\mathbf{S}}_{i+1} \\
+ \frac{J}{2} \sum_{i}  \hat{\mathbf{S}}_i \cdot \hat{\mathbf{S}}_{i+2} - J \sum_j \hd_j \h_j \hat{\mathbf{S}}_j \cdot \hat{\mathbf{S}}_{j+1}.
\label{eq:MGmodel}
\end{multline}
Note that the impurity is only switching off nearest neighbor interactions in our toy model. We defined the spectral function as usual, by removing a spin and creating the hole-like impurity at the same site.

Without the hole, the Majumdar-Ghosh model describes certain materials, where the atoms form a zig-zag chain \cite{Lacroix2011}, see Fig.~\ref{figxMG} a), and constitutes an example of an exactly solvable, frustrated spin system. Its degenerate ground states are exactly known, for a pedagogical discussion see e.g. Ref.~\cite{Auerbach1998}. They spontaneously break the translational symmetry and consist of states in which neighboring spins form a singlet. The two lowest energy states with this property, which are related to each other by a shift by one lattice site, are degenerate in the thermodynamic limit and are separated from the excitation spectrum by a gap.

In Fig.~\ref{figxMG} b), the spectral function of a single hole in a Majumdar-Ghosh spin chain is shown. Gray dots correspond to the combined dispersion relation of the holon and the spinon, $\epsilon(k)=\epsilon_h(k_h) + \epsilon_s(k-k_h)$ with $\epsilon_h(k_h)=-2t\cos k_h$ and the Majumdar-Ghosh dispersion relation~\cite{Shastry1981} 
\begin{align}
\epsilon_s(k_s) &= J\frac{5}{4}+\frac{J}{2} \cos (2 k_s),
\label{eq:spinon_dispersion_MG}
\end{align}
where $k_s$ is restricted to half of the Brillouin zone. The distinct lines in the spectrum are remarkably well described by Eq. \eqref{eq:spinon_dispersion_MG}, demonstrating that spin-charge separation applies.
 
The comparison to the isotropic Heisenberg spin chain without frustration highlights an interesting feature. An asymmetry in the distribution of spectral weight around $|k-k_h| = \pi/2$ is clearly visible. However, the spectral weight is not as clearly restricted to half of the Brillouin zone as in the spectrum of the Heisenberg chain. Thus, a valence bond solid nature of the ground state is not sufficient to explain the sharp decrease of spectral weight observed for the anisotropic spin chain in Fig.~\ref{figIsing}.

\section{Squeezed space formalism for one hole in a spin chain}
\label{AppdxSBPderivation}
In this appendix we derive the effective Hamiltonian describing a single hole inside a one-dimensional spin chain. For concreteness we discuss the $t-J^*$ Hamiltonian from Eq.~\eqref{eq:tjmodel}, but generalizations to other couplings are straightforward. The hole can be described by a bosonic representation where the spins are mapped to constrained fermions $\f_{j,\sigma}$ and the holons to bosonic operators $\hd_j$. In this case $\c_{j,\sigma} = \hd_j \f_{j,\sigma}$. This representation was used in our theoretical analysis in Sec.~\ref{secTheory} and we discuss it here in more detail.

After introducing bosonic operators $\h_j$ and spinons $\f_{j,\sigma}$ as discussed in the beginning of Sec.~\ref{secTheory}, we can simplify the holon degree of freedom by effectively removing it from the spin chain. This can be achieved by defining a basis of the Hilbert space of a spin chain with a single hole, with basis states $\hd_j \ket{0} \otimes \ket{\tilde{\sigma}_1,...,\tilde{\sigma}_{L-1}}$, where
\begin{equation}
\ket{\tilde{\sigma}_1,...,\tilde{\sigma}_{L-1}} \equiv  \tfd_{1,\tilde{\sigma}_1} ... \tfd_{L-1,\tilde{\sigma}_{L-1}} \ket{0}
 \label{eqSigmaDefFSq}
\end{equation}
similarly to Eq.~\eqref{eqSigmaDefF}. Here, $\tilde{\sigma}_i={\uparrow}$, $\downarrow$. The index $i$ labels spins in the chain from left to right, independent of the holon position. Note that only $L-1$ spins appear in Eq.~\eqref{eqSigmaDefFSq} because there is no spin on the physical site occupied by the holon, and we used the spinon operators $\tf$ to define a constrained fermion representation of spins in squeezed space.

Next we need to express the original Hamiltonian, Eq.~\eqref{eq:tjmodel}, formulated using operators $\c_{j,\sigma}$, in the new basis. Without doping, the Hamiltonian 
\begin{equation}
\H = J \sum_{j} \hat{\mathbf{S}}_{j+1} \cdot \hat{\mathbf{S}}_j, \qquad \hat{\mathbf{S}}_i = \frac{1}{2} \f_{i,\alpha}^\dagger \boldsymbol{\sigma}_{\alpha,\beta} \f_{i,\beta}
\end{equation}
corresponds to a Heisenberg spin chain. In the case with doping, there exist no exchange interactions between the two spins adjacent to the hole. For a single holon we can thus write the Hamiltonian in Eq.~\eqref{eq:tjmodel} as $\H_{t-J^*} = \H_t+ \H_J+\H_{\rm{NNN}}$ with
\begin{equation}
\H_J = J \sum_{j} \tilde{\mathbf{S}}_{j+1} \cdot \tilde{\mathbf{S}}_j \l 1 - \hd_j \h_j \r , \quad \tilde{\mathbf{S}}_i = \frac{1}{2} \tf_{i,\alpha}^\dagger \boldsymbol{\sigma}_{\alpha,\beta} \tf_{i,\beta}.
\label{eqHintJsqueezed}
\end{equation}
The hopping term
\begin{equation}
\H_t =  -t ~ \mathcal{P} \sum_{\ij, \sigma} \cd_{i,\sigma} \c_{j,\sigma}  \mathcal{P},
\end{equation}
moves the holon by one site while the order of fermions $\tf$ in squeezed space is not modified. One can write $\H_t$ most conveniently as
\begin{equation}
\H_t = - t \sum_{\ij} \hd_{j} \h_i = - 2 t \sum_k \cos (k) \hd_k \h_k. 
\end{equation}

For the next-nearest neighbor tunnelings, the situation is more complicated. They are of a general form
\begin{equation}
\H_{\rm NNN} = \sum_{\sigma,\sigma',\tau,\tau'} g_{\sigma,\sigma',\tau,\tau'} \sum_i \cd_{i+2,\sigma}  \cd_{i+1,\tau} \c_{i+1,\tau'} \c_{i,\sigma'} + \hc, 
\label{eq:HNNNgen}
\end{equation}
where the coefficients $g$ can be read off from Eq.~\eqref{eq:tjmodel}.
The term in Eq.~\eqref{eq:HNNNgen} modifies the order of spins because it moves a fermion from site $i$ to $i+2$. This involves an exchange of the fermions at sites $i$ and $i+1$, which introduces an additional minus sign. We can see this by calculating the action of $\H_{\rm NNN}$ on a basis state:
\begin{equation}
\H_{\rm NNN} \hd_{i+2} \ket{0} \otimes \ket{...,\tilde{\sigma}_{i},\tilde{\sigma}_{i+1},...} .
\end{equation}
To lighten the notation, we consider the action of a single term in the sum in Eq.~\eqref{eq:HNNNgen}. Representing the basis state in terms of the original $\cd_{i,\sigma}$ operators yields
\begin{multline}
\cd_{i+2,\sigma}  \cd_{i+1,\tau} \c_{i+1,\tau'} \c_{i,\sigma'} \cd_{1,\tilde{\sigma}_1} ... \cd_{i,\tilde{\sigma}_i} \cd_{i+1,\tilde{\sigma}_{i+1}} ... \ket{0} \\
=  \cd_{1,\tilde{\sigma}_1} ...  \cd_{i-1,\tilde{\sigma}_{i-1}} (-1)^{i-1} \delta_{\sigma', \tilde{\sigma}_i} \delta_{\tau',\tilde{\sigma}_{i+1}}\\
\times  \cd_{i+1,\tau} (-1)^i \cd_{i+2,\sigma} ... \ket{0}.
\end{multline}
In the notation introduced above, this state can be identified with
\begin{equation}
 - \delta_{\sigma', \tilde{\sigma}_i} \delta_{\tau',\tilde{\sigma}_{i+1}}  \hd_{i+2}  \otimes \ket{...,\tilde{\sigma}_{i-1},\tau,\sigma,\tilde{\sigma}_{i+2},...} .
\label{eqSqueezedSpaceMinusSign}
\end{equation}
The next-nearest neighbor term leads to an exchange of the spins on sites $i$ and $i+1$ in squeezed space which can be described by a term of the form $4 \tilde{\vec{S}}_i \cdot \tilde{\vec{S}}_{i+1} - 1$, see e.g. supplementary material in Ref.~\cite{Hilker2017}. By taking into account the minus sign from the fermion exchange above, we arrive at the following expression,
\begin{equation}
\H_{\rm NNN} = \frac{J}{2}  \sum_{j} \hd_{j+2} \h_j \l \tilde{\vec{S}}_{j+1} \cdot \tilde{\vec{S}}_{j} - \frac{1}{4} \r + \hc.
\label{eqHNNNsqueezedBosonh}
\end{equation}
From the second term in the brackets, we derive the corresponding term $-\frac{1}{4}J\cos(2k_h)$ in the holon dispersion, Eq.~\eqref{eq:holon_dispersion}. 

In summary, the $t-J^*$ Hamiltonian for a single hole can be written in squeezed space as
\begin{multline}
\H_{t-J^*} = - \sum_{k_h} \hd_{k_h} \h_{k_h} \l 2 t \cos (k_h)  + \frac{1}{4}J\cos(2k_h)  \r\\
+  \frac{J}{2}  \sum_{j} \left[ \hd_{j+2} \h_j  \tilde{\vec{S}}_{j+1} \cdot \tilde{\vec{S}}_{j} + \hc \right]\\
+ J \sum_{j} \tilde{\mathbf{S}}_{j+1} \cdot \tilde{\mathbf{S}}_j \l 1 - \hd_j \h_j \r.
\label{eqEffictiveTeJayStar}
\end{multline}

\section{Effective polaron description}
\label{secPolaronDescription}
We study the interaction of a single holon with the surrounding spin environment. It can be created by first removing a fermion from the spin chain, leading to the creation of a spinon-holon pair. The essence of spin-charge separation is that the spinon is not bound to the holon. Thus, after waiting for sufficiently long, we can assume that spinon and holon propagate through the system independently. We now provide a detailed calculation of the holon properties when it becomes dressed by collective spin excitations and forms a holon-polaron. In particular, we derive Eqs. \eqref{eq:EhPiComp}, \eqref{eq:MhPiComp} from the main text for its renormalized energy and the renormalized mass.

Starting from the $t-J^*$ Hamiltonian, we first perform a Lee-Low-Pines transformation into the holon-frame. As a next step, bosonization techniques are employed to describe the spin chain itself, as well as the interaction of the holon with the collective spin excitations. Here we assume an infinite system and neglect finite-size corrections. Finally, we explicitly calculate the holon energy and its renormalized mass in the regime of weak polaronic coupling, $J \ll t$.

\subsection{Lee-Low-Pines transformation}\label{secLLP}
Our starting point is the $t-J^*$ Hamiltonian \eqref{eq:tjmodel} formulated in squeezed space \cite{Kruis2004a}, see Eq.~\eqref{eqEffictiveTeJayStar}, where the spinless holon is effectively hopping between the bonds of the physical lattice. As a first step, we apply the unitary Lee-Low-Pines transformation \cite{Lee1953}, 
\begin{equation}
\hat{U}_{\rm LLP} = \exp \l i \hat{P}_{\rm s} \hat{x}_{\rm h} \r
\end{equation}
where $\hat{x}_{\rm h} = \sum_j j \hd_j \h_j$ is the position operator of the holon and $\hat{P}_{\rm s}$ is the total momentum operator of the spins. In the new basis the holon is always placed in the center, and the transformed Hamiltonian reads
\begin{multline}
\hat{U}_{\rm LLP}^\dagger \H \hat{U}_{\rm LLP} = - 2 t \cos \l p_{\rm h} -\hat{P}_{\rm s} \r + J \sum_\ij \hat{\vec{S}}_i \cdot  \hat{\vec{S}}_j \\
- J \hat{\vec{S}}_0 \cdot  \hat{\vec{S}}_{1} - \frac{J}{4} \cos \l 2 p_{\rm h} - 2 \hat{P}_{\rm s} \r \\
+ \frac{J}{2} \biggl( \hat{\vec{S}}_{-1} \cdot \hat{\vec{S}}_0 e^{2 i (p_{\rm h} - \hat{P}_{\rm s})} + \hat{\vec{S}}_{1} \cdot \hat{\vec{S}}_2 e^{- 2 i (p_{\rm h} - \hat{P}_{\rm s})} \biggr) 
\label{eq:HamHolonPolaronLLP}
\end{multline}
Here $p_{\rm h}$ denotes the total conserved momentum of the holon-polaron. 

The first term in Eq.~\eqref{eq:HamHolonPolaronLLP} corresponds to the recoil energy of the holon when it scatters on a spin-wave excitation which changes the total momentum $\hat{P}_{\rm s}$ carried by the spin system. The second term describes the unperturbed spin chain without the holon. The interaction between the holon and the spin chain within the $t-J$ model is given by $- J \hat{\vec{S}}_0 \cdot \hat{\vec{S}}_1$. The last three terms in the equation describe next-nearest neighbor hopping processes present in the Fermi-Hubbard model at large $U$. These terms have the same scaling with $J$ as the interactions $- J \hat{\vec{S}}_0 \cdot \hat{\vec{S}}_1$ in the simpler $t-J$ model and should be treated on equal footing to understand the properties of the holon-polaron. 

\subsection{Bosonization}
\begin{figure}[t!]
\centering
\includegraphics[width=0.28\textwidth]{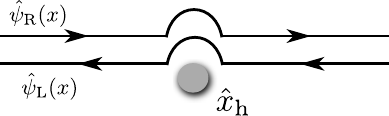}
\caption{\textbf{Bosonization of a spin chain doped by a mobile holon.} The spin operators can be described by two chiral fermions $\hat{\psi}_{\rm L,R}(x)$ interacting with the holon at position $\hat{x}_{\rm h}$. Our starting point is the unperturbed chain, which is an exact solution when the tunneling rate $t$ of the holon is larger than spin-exchange interactions, $t \gg J$.}
\label{figBosonization}
\end{figure}

In order to calculate ground state properties of Hamiltonian \eqref{eq:HamHolonPolaronLLP}, we use the bosonization technique to describe the unperturbed spin chain. Here we provide a brief overview, see e.g. Refs.~\cite{Haldane1981,Giamarchi2003,Eggert2007} for a complete derivation. In this formalism, spin operators are first expressed in terms of Jordan-Wigner fermions $\ps(x)$ counting the number of up spins. In this basis $\hat{S}^z(x) = \psd(x) \ps(x)-1/2$ and the case of zero magnetization corresponds to a system at half filling. When the net magnetization vanishes and mutual interactions between the Jordan-Wigner fermions are neglected, they form a Fermi sea around $k_{\rm F}=\pi/2$, in units of the inverse lattice constant $a=1$. By linearizing around $k_{\rm F}$ and extending the linearized branches to momenta $\pm \infty$, one can introduce two chiral fields $\ps_{\rm R,L}(x)$ corresponding to right (R) and left (L) movers. The situation is illustrated in Fig.~\ref{figBosonization}. 

The interactions between Jordan-Wigner fermions renormalize the properties of collective particle-hole excitations around the Fermi surface. In the bosonization formalism this is described by introducing bosonic fields $\hat{\phi}_{\rm L,R}(x)$, related to the Jordan-Wigner fermions by $\ps_{\rm L,R}(x) \propto \exp (\mp i \sqrt{4 \pi} \hat{\phi}_{\rm L,R}(x))$. The fields $\hat{\phi}_{\rm L,R}(x)$ can be decomposed into normal modes $\b_q$ with momentum $q >0$ for right-movers ($q <0$ for left-movers), where $[\b_q,\bd_{q'}]=\delta_{q,q'}$. In addition, so-called zero modes have to be included, where $\hat{n}_{\rm L,R}$ counts the total number of left and right movers relative to the Fermi sea at $k_{\rm F}=\pm \pi/2$. Their number can be changed by the operators $\exp( i \sqrt{4\pi} \hat{\phi}_0^{\rm L,R})$, namely $\exp( i \sqrt{4\pi} \hat{\phi}_0^{\rm L,R}) \hat{n}_{\rm L,R} = (\hat{n}_{\rm L,R} \mp 1) \exp( i \sqrt{4\pi} \hat{\phi}_0^{\rm L,R})$. Putting everything together one obtains \cite{Eggert2007}
\begin{equation}
 \hat{\phi}_{\rm L,R}(x) = \hat{\phi}_0^{\rm L,R} + \frac{\sqrt{\pi} x}{L} \hat{n}_{\rm L,R} + \sum_{q \lessgtr 0} \frac{{\rm sgn}(q)}{\sqrt{2 L |q|}} \l e^{i q x} \b_q + \hc \r,
\end{equation}
where $L$ is the length of the spin chain (not counting the holon) and we assumed periodic boundary conditions. We furthermore introduce the shorthand notation $ \hat{\phi}_{\rm L,R}(x) = \hat{\phi}_0^{\rm L,R} + \frac{\sqrt{\pi} x}{L} \hat{n}_{\rm L,R} + \hat{\phi}_{\rm L,R}^\lessgtr(x)$.

Using the formalism described above, the collective excitations of the anti-ferromagnetic spin chain can be described as
\begin{flalign}
\H_0 &= J \sum_\ij \hat{\vec{S}}_i \cdot  \hat{\vec{S}}_j \\
&=  E_0 + \pi J \int dx :  \bigl( \partial_x \hat{\phi}(x) \bigr)^2 + \frac{1}{4}  \bigl( \partial_x \hat{\theta}(x) \bigr)^2  :   \label{eq:HLLLintegral} \\
&= E_0 + \sum_q v_{\rm s} |q| \bd_q \b_q + J\frac{\pi}{L}\left( \hat{n}_{\rm L}^2 +\hat{n}_{\rm R}^2 \right).
\end{flalign}
In the second line we introduced $\hat{\phi}(x) = \hat{\phi}_{\rm R}(x) + \hat{\phi}_{\rm L}(x)$, $\hat{\theta}(x) = \hat{\phi}_{\rm R}(x) - \hat{\phi}_{\rm L}(x)$ and $:...:$ denotes normal ordering with respect to operators $\b_q$. Note that Umklapp terms have been neglected in this expression. The ground state energy $E_0$ of the spin chain is known exactly from Bethe-ansatz calculations \cite{Giamarchi2003},
\begin{equation}
E_0 = L \l \nicefrac{1}{4} - \log(2) \r J = - L \times 0.4431 ~ J.
\label{eq:E0LLL}
\end{equation}
The excitations described by $\bd_q$ carry no spin and their velocity $v_{\rm s}$ is given by \cite{Giamarchi2003}
\begin{equation}
v_{\rm s} = J \pi/2.
\end{equation}
The ultraviolet momentum cut-off $\Lambda_{\rm UV}$ is determined by the inverse lattice scale $1/a$. We set $\Lambda_{\rm UV} = \pi/2$ in units where $a=1$. 

The total momentum operator of the spin system contains contributions from collective excitations as well as the zero modes. Assuming that $|n_{\rm L,R}| \ll L$ for typical values of $\hat{n}_{\rm L,R}$, we can write
\begin{equation}
\hat{P}_{\rm s} = \frac{\pi}{2} \l \hat{n}_{\rm R} - \hat{n}_{\rm L} \r + \sum_q q \bd_q \b_q.
\label{eq:Pstot}
\end{equation}

\subsection{Interaction terms}
The interaction of the holon with collective spin excitations is determined by the term
\begin{equation}
\H_{\rm int} = - J \hat{\vec{S}}_0 \cdot \hat{\vec{S}}_1 + \frac{J}{2} \l \hat{\vec{S}}_{-1} \cdot \hat{\vec{S}}_0 e^{2 i (p_{\rm h} - \hat{P}_{\rm s})} + \hc \r
\label{eq:Hint}
\end{equation}
in the Lee-Low-Pines frame, see Eq.~\eqref{eq:HamHolonPolaronLLP}. To express it in terms of bosonized operators, we first note that the energy density of the free spin chain is given by $J \hat{\vec{S}}(x) \cdot \hat{\vec{S}}(x+1)$. By inspection of Eq.~\eqref{eq:HLLLintegral} we can write,
\begin{equation}
\hat{\vec{S}}(x) \cdot \hat{\vec{S}}(x+1) \approx \frac{E_0}{J L} + :  \pi  \bigl( \partial_x \hat{\phi}(x) \bigr)^2 + \frac{\pi }{4}  \bigl( \partial_x \hat{\theta}(x) \bigr)^2 :~.
\end{equation}
This term describes correctly the effect of long-wavelength fluctuations on nearest neighbor spin correlations. Because of the constant term added to the expression, the ground state expectation value $\langle \hat{\vec{S}}(x) \cdot \hat{\vec{S}}(x+1) \rangle$, is correctly reproduced.

In combination with Eq.~\eqref{eq:E0LLL} we obtain the following representation,
\begin{multline}
\hat{\vec{S}}(x) \cdot \hat{\vec{S}}(x+1) = \nicefrac{1}{4} - \log(2) +  \pi^2 \l \frac{\hat{n}_-^2}{4 L^2} + \frac{\hat{n}_+^2}{L^2} \r \\
+ \pi \l \frac{\hat{n}_+}{L} + \frac{\hat{n}_-}{4 L} \r    \sqrt{\frac{2 \pi}{L}} \sum_q \sqrt{|q|} e^{i q x} i \l \b_q - \bd_{-q}  \r \\
+ \frac{\pi}{4 L} \sum_{qq'>0} \sqrt{|qq'|}~ :  \underline{\hat{b}}^\dagger_q W_{q,q'}(x)  \underline{\hat{b}}_{q'} : \\
+ 4 |C|^2 e^{i \pi x}  \cos \left[ \sqrt{4 \pi} ~ \l \hat{\phi}_{\rm L}(x)  + \hat{\phi}_{\rm R}(x) \r \right].
\label{eq:SdotSbos}
\end{multline}
Here, $\hat{n}_\pm=\hat{n}_{\rm R} \pm \hat{n}_{\rm L}$ and we defined
\begin{equation}
W_{q,q'}(x) = \l 
\begin{array}{c c}
e^{- i (q-q')x} & -e^{- i (q+q')x} \\
-e^{ i (q+q')x} & e^{ i (q-q')x}
\end{array}
\r
\end{equation}
and $\underline{\hat{b}}^\dagger_q = (\bd_q, \b_q)$. In the last line of Eq.~\eqref{eq:SdotSbos} we have also included Umklapp scattering terms. Their prefactor is the number $|C|^2$, which is a non-universal constant that cannot be derived within the bosonization formalism. Its numerical value $|C|^2\approx 0.14$ has been determined by Eggert and Affleck \cite{Eggert1995} from comparison of the spin-structure factor obtained from bosonization and quantum Monte Carlo calculations.

The first three lines in Eq.~\eqref{eq:SdotSbos} describe forward scattering processes, which leave the populations $\hat{n}_{\rm L,R}$ of the zero-modes unchanged. The last line corresponds to Umklapp scattering, where the $\hat{n}_-$ changes by two units because a right mover-scatters into a left-mover or vice-versa. The sum $\hat{n}_+ = \hat{n}_{\rm R} + \hat{n}_{\rm L} = m L$ is conserved and can be expressed in terms of the magnetization per unit length 
\begin{equation}
m=(N_\uparrow-N_\downarrow)/2L.
\end{equation}
In the following we allow for a finite magnetization $m \neq 0$ but assume that $\hat{n}_- = \mathcal{O}(L^0)$ is not an extensive quantity. 

To understand the scaling in the thermodynamic limit $L \to \infty$, we introduce operators $\b(q) = \sqrt{L / 2 \pi} \b_q$ with $[\b(q),\bd(q')] = \delta(q-q')$ and write Eq.~\eqref{eq:SdotSbos} in an integral form using $\sum_q 2 \pi/L \to \int dq$ with $q$ integrated over $-\Lambda_{\rm UV} \leq q \leq \Lambda_{\rm UV}$,
\begin{multline}
\hat{\vec{S}}(x) \cdot \hat{\vec{S}}(x+1) = \nicefrac{1}{4} - \log(2) + \pi^2 m^2 \\
+ \pi m \int dq~\sqrt{|q|} e^{i q x}  i \l \b(q) - \bd(-q)  \r  \\
 + \frac{1}{8} \int_{qq'>0} dqdq'~ \sqrt{|qq'|} ~ : \underline{\hat{b}}^\dagger(q) W_{q,q'}(x)  \underline{\hat{b}}(q') :\\
+ 4 |C|^2 e^{i \pi x}  \cos \left[ \sqrt{4 \pi} ~ \l \hat{\phi}_{\rm L}(x)  + \hat{\phi}_{\rm R}(x) \r \right]  .
\label{eq:SdotSbosInt}
\end{multline}
The field $\hat{\phi}_{\rm R}(x)$ can be expressed as
\begin{equation}
 \hat{\phi}_{\rm R}(x) = \hat{\phi}_0^{\rm R} + \frac{\sqrt{\pi} x}{L} \hat{n}_{\rm R} + \int_0^{\Lambda_{\rm UV}} dq~  \frac{|q|^{-1/2}}{\sqrt{4 \pi}} \l e^{i q x} \b(q) + \hc \r,
\end{equation}
and a similar expression exists for $\hat{\phi}_{\rm L}(x)$.

The interactions in Eq.~\eqref{eq:Hint} can now be written in the bosonization language using Eq.~\eqref{eq:SdotSbosInt}. We will distinguish forward scattering terms (F) of different orders in $\b$ and Umklapp (U) scattering terms,
\begin{multline}
\H_{\rm int} =  \H_{\rm F}^{(1)} + \H_{\rm F}^{(2)} + \H_{\rm U}+ \l  \nicefrac{1}{4} - \log(2) + \pi^2 m^2 \r \\
\times J  \left[ \cos \l 2 (p_{\rm h} - \hat{P}_{\rm s}) \r - 1 \right].
\label{eqHintHF1HF2add}
\end{multline}

\subsubsection{Linear Fr\"ohlich-type terms}

First we consider only forward scattering terms that are linear in $\b$ operators in Eq.~\eqref{eq:Hint},
\begin{multline}
\H_{\rm F}^{(1)} = J \pi m \int dq~\sqrt{|q|}   i \l \b(q) - \bd(-q)  \r \\ 
\times \left[ 1 - \cos \l 2 (p_{\rm h} - \hat{P}_{\rm s})- q \r \right]
\label{eq:Hfrohlich}
\end{multline}
Notably, the form of this term is identical to the interaction terms in the ubiquitous Fr\"ohlich polaron Hamiltonian \cite{Froehlich1954,Catani2012,Grusdt2016RG1D}. 

This allows us to introduce a dimensionless polaronic coupling constant $\alpha_{\rm F} = (\pi m)^2$. When $\alpha_{\rm F} \ll 1$, the Fr\"ohlich coupling is weak and can be treated perturbatively. The starting point for such analysis is Eq.~\eqref{eqHintHF1HF2add} with $ \H_{\rm F}^{(1,2)} = 0$ when $\alpha_{\rm F}=0$. The resulting Hamiltonian commutes with $\bd_q \b_q$ and can be solved exactly by plane waves. When $\alpha_{\rm F} \gg 1$, on the other hand, the strong-coupling (or Landau-Pekar) variational wavefunction can be used \cite{Landau1948,Casteels2011}. Note that our derivation is only valid at weak polaron couplings $\alpha_{\rm F} \ll 1$. By expanding around the Fermi-sea at half filling of Jordan-Wigner fermions, we assumed that $m \ll 1$ is small.

\subsubsection{Two-particle excitations}
By keeping only forward scattering terms which are quadratic in $\b$ operators in Eq.~\eqref{eq:Hint}, we obtain
\begin{multline}
\H_{\rm F}^{(2)} = - \frac{J}{8} \int_{qq'>0} dqdq'~ \sqrt{|qq'|} ~ : \biggl[ \underline{\hat{b}}^\dagger(q) W_{q,q'}(0)  \underline{\hat{b}}(q') \\
- \frac{1}{2} \underline{\hat{b}}^\dagger(q) \l \tilde{W}_{q,q'}(-1) e^{2 i (p_{\rm h} - \hat{P}_{\rm s})} + \hc \r \underline{\hat{b}}(q')  \biggr] :.
\label{eq:HtwoPart}
\end{multline}
The matrix $W_{q,q'}(x)$ in the LLP frame becomes
\begin{equation}
\tilde{W}_{q,q'}(x) = \l 
\begin{array}{c c}
e^{- i (q-q')x}  e^{-2i q'} & -e^{- i (q+q')x}  e^{2i q'}\\
-e^{ i (q+q')x} e^{-2i q'} & e^{ i (q-q')x} e^{2i q'}
\end{array}
\r
\end{equation}
For a system at zero magnetization, $m=0$, these are the leading-order terms in the effective holon-polaron Hamiltonian.

\subsubsection{Umklapp scattering}
Finally, Umklapp scattering terms give rise to the following interactions,
\begin{multline}
\H_{\rm U} = - 2 J |C|^2 \biggl[2 \cos \l \sqrt{4 \pi} \hat{\phi}(0) \r  + \cos \l \sqrt{4 \pi} \hat{\phi}(-1) \r \\ 
\times e^{2 i (p_{\rm h} - \hat{P}_{\rm s})} + \cos \l \sqrt{4 \pi} \hat{\phi}(1) \r e^{-2 i (p_{\rm h} - \hat{P}_{\rm s})}  \biggr],
\label{eq:Humklapp}
\end{multline}
where $\hat{\phi}(x) = \hat{\phi}_{\rm L}(x) +\hat{\phi}_{\rm R}(x)$.

\subsection{Holon polaron at weak polaronic coupling}
Now we analyze the properties of the holon polaron in the weak polaronic coupling regime, $J \ll t$. Our starting point is a free holon at momentum $p_{\rm h}$ and no spin excitations,
\begin{equation}
\ket{\psi_0} = \ket{p_{\rm h}}_h ~ \ket{0}_0 ~ \ket{0}_b.
\end{equation}
The occupation of the zero modes is characterized by $\ket{n_-}_0$ and $n_+ = L m$ is conserved.

To zeroth order the holon energy is given by
\begin{equation}
E_{\rm h}^{(0)} = - 2 t \cos (p_{\rm h}).
\label{eq:Efwd0}
\end{equation}
We obtain the following first-order contribution to the holon-polaron energy,
\begin{equation}
E_{\rm h}^{(1)} =  J \l  \pi^2 m^2 - \log(2)  \r \left[ \cos \l 2 p_{\rm h} \r - 1 \right] - J/4.
 \label{eq:Efwd1}
\end{equation}

In the following we calculate second order corrections in $J$ to the holon-polaron energy term by term. Furthermore, by expanding the result around $p_{\rm h} =0$ to quadratic order, we calculate the mass renormalization of the holon-polaron. 

\subsubsection{Forward scattering}
From the Fr\"ohlich-type terms \eqref{eq:Hfrohlich} we obtain the following second-order contribution,
\begin{equation}
E_{\rm h, F1}^{(2)} = - J^2 \pi^2 m^2 \int dq~|q| \frac{\left[ 1 - \cos (2 p_{\rm h} - q ) \right]^2}{2 t \cos(p_{\rm h})-2 t \cos(p_{\rm h} - q)}.
\label{eq:EfwdF1}
\end{equation}
From the two-particle terms \eqref{eq:HtwoPart} we obtain a second-order contribution
\begin{multline}
E_{\rm h, F2}^{(2)} = - \frac{J^2}{64} \int_{qq'>0} dqdq'~ |qq'| \frac{\left[ 1 - \cos \l 2 p_{\rm h} + q + q' \r \right]^2}{\Omega_{q,q'}(p_{\rm h})},
\label{eq:EfwdF2}
\end{multline}
where $\Omega_{q,q'}(p_{\rm h})=-2 t \cos(p_{\rm h} - q-q') + 2 t \cos(p_{\rm h})$.

\subsubsection{Umklapp scattering}
Now we calculate the leading-order contribution of the Umklapp scattering term Eq.~\eqref{eq:Humklapp} to the holon-polaron energy. We start by noting that
\begin{equation}
e^{\pm i \sqrt{4 \pi} \left(\hat{\phi}_0^{\rm L}+\hat{\phi}_0^{\rm R} + \frac{\sqrt{\pi} x}{L} (\hat{n}_{\rm L}+\hat{n}_{\rm R})\right)}\ket{n_-}_0 = \ket{n_- \pm 2}_0.
\end{equation}
The momentum is given by $\hat{P}_{\rm s} \ket{n_-}_0 = n_- \pi/2$, see Eq.~\eqref{eq:Pstot}. The action of $\exp ( i \sqrt{4\pi} \hat{\phi}(x) )$ on the bosonic state $\ket{0}_b$ can be understood by writing
\begin{equation}
e^{i \sqrt{ 4 \pi}\hat{\phi}_b(x) } \ket{0}_b = \exp \l - \int dq ~ \beta_x^*(q) \b(q) - \hc  \r \ket{0}_b,
\end{equation}
where $\hat{\phi}_b(x) = \hat{\phi}_{\rm L}^<(x)+\hat{\phi}_{\rm R}^>(x)$. The last expression corresponds to a coherent state $\prod_q \ket{\beta_x(q)}$, with 
\begin{equation}
\beta_x(q) = i ~ {\rm sgn}(q) ~ |q|^{-1/2} e^{- i q x}.
\end{equation}
Note that the amplitude
\begin{equation}
|\beta_x(q)|^2 = |q|^{-1} \equiv |\beta(q)|^2
\end{equation}
is independent of $x$. 

By summing over the allowed virtual states, defined by Fock states of $\b$ operators and $\ket{\pm 2}_0$, we obtain an expression for the holon-polaron energy due to Umklapp scattering,
\begin{multline}
E_{\rm h, U}^{(2)} = 8 J^2 |C|^4 \sum_{n_q} \l \prod_q  \frac{|\beta(q)|^{2 n_q}}{n_q!}e^{- |\beta(q)|^2} \r \frac{1}{\Omega_Q(p_{\rm h})} \\
\times \l  \frac{3}{2} + \frac{1}{2} \cos (2 Q + 4 p_{\rm h}) + 2 \cos (Q + 2 p_{\rm h})\r.
\label{eq:Eumklapp}
\end{multline}
The energy denominator is given by
\begin{equation}
\Omega_Q(p_{\rm h}) = - 2 t \cos(p_{\rm h}) + 2t \cos(p_{\rm h} - \pi - Q),
\end{equation}
and depends only on the total momentum 
\begin{equation}
Q = \int dq~ q ~ n_q.
\end{equation}

To simplify Eq.~\eqref{eq:Eumklapp} we note that the sum $\sum_{n_q}$ of the occupations $n_q$ is taken over
\begin{equation}
\lambda(n_q) = \frac{|\beta(q)|^{2 n_q}}{n_q!}e^{- |\beta(q)|^2}.
\end{equation}
This defines independent Poisson distributions for all momentum modes $q$. Because the remaining terms only depend on the total momentum $Q$, it is sufficient to know the full counting statistics $p(Q)$ of the latter with respect to the independent Poisson distributions:
\begin{multline}
E_{\rm h, U}^{(2)} = 8 J^2 |C|^4 \sum_{Q} p(Q) \frac{1}{\Omega_Q(p_{\rm h})} \\
\times \l  \frac{3}{2} + \frac{1}{2} \cos (2 Q + 4 p_{\rm h}) + 2 \cos (Q + 2 p_{\rm h})\r.
\label{eq:EumklappFCS}
\end{multline}

To calculate the full counting statistics $p(Q)$, we construct the generating functional
\begin{multline}
G(X) = \sum_{n_q} \lambda(n_q) e^{- i X Q} \\
= \exp \left[ - \int dq ~ |\beta(q)|^2 \l 1 - e^{- i X q} \r \right] \\
= e^{- 2 \gamma} \frac{4}{\pi^2 X^2} e^{2 {\rm Ci} (X \pi/2)}.
\end{multline}
Here $\gamma=0.577216$ is the Euler constant and ${\rm Ci}(x)=-\int_x^\infty ~ \cos(t)/t~dt$ denotes the cosine integral. 
By taking a Fourier transform we obtain the full counting statistics,
\begin{equation}
p(Q) = \frac{1}{2 \pi} \int dX ~ G(X) e^{i X Q}.
\end{equation}
Combining this with Eq.~\eqref{eq:EumklappFCS} we arrive at
\begin{multline}
E_{\rm h, U}^{(2)} = J^2 |C|^4 e^{-2 \gamma} \frac{2^5}{\pi^3} \int_{-\infty}^\infty dQ \int_0^\infty dX~\frac{\cos ( X Q) }{X^2 \Omega_Q(p_{\rm h})} \\
\times e^{2 {\rm Ci} (X \pi/2)} \l  \frac{3}{2} + \frac{1}{2} \cos (2 Q + 4 p_{\rm h}) + 2 \cos (Q + 2 p_{\rm h})\r.
\label{eq:EumklappFCScompl}
\end{multline}

\subsubsection{Renormalized mass and energy around $p_{\rm h}=0$}
Now we analyze the results from Eqs.~\eqref{eq:Efwd0}-\eqref{eq:EfwdF2}, \eqref{eq:EumklappFCScompl} and calculate the holon-polaron ground state properties at $p_{\rm h}=0$. Up to quadratic order in $p_{\rm h}$ we obtain
\begin{equation}
E_{\rm h}(p_{\rm h}) = E_{\rm h}(0) + \frac{1}{2} p_{\rm h}^2 M_{\rm h}^{-1} + \mathcal{O}(p_{\rm h}^4).
\end{equation}
The ground state energy $E_{\rm h}(0)$ contains contributions from the five different terms in the effective Hamiltonian,
\begin{equation}
E_{\rm h}(0) = E_{\rm h}^{(0)}(0) + E_{\rm h}^{(1)}(0) + E_{\rm h, F1}^{(2)}(0) + E_{\rm h, F2}^{(2)}(0) + E_{\rm h, U}^{(2)}(0).
\end{equation}
A similar expression follows for the holon-polaron mass,
\begin{equation}
\frac{1}{M_{\rm h}} = \frac{1}{M_{\rm h}^{(0)}} + \frac{1}{M_{\rm h}^{(1)}} + \frac{1}{M_{\rm h, F1}^{(2)} } +\frac{1}{M_{\rm h, F2}^{(2)}} + \frac{1}{M_{\rm h, U}^{(2)} }.
\end{equation}

The different contributions are given by
\begin{flalign}
 E_{\rm h}^{(0)}(0) &= -2 t \\
 E_{\rm h}^{(1)}(0)  &= - J/4 \\
 E_{\rm h, F1}^{(2)}(0)  &= - \frac{J^2}{t} m^2 \pi^2  \l 1 + \frac{\pi^2}{8} - \frac{\pi}{2} \r \\
 E_{\rm h, F2}^{(2)}(0)  &= - \frac{J^2}{t} \frac{\pi}{4096} (64 - 16 \pi + \pi^3) \\
 E_{\rm h, U}^{(2)}(0) &= - \frac{J^2}{t} |C|^4 \times 5.31.
\end{flalign}
By combining these results, we obtain Eq.~\eqref{eq:EhPiComp}. For the effective mass we obtain
\begin{flalign}
 \l M_{\rm h}^{(0)} \r^{-1} &= 2 t \\
 \l M_{\rm h}^{(1)} \r^{-1} &= - 4 J \left[ \log(2) -\pi^2 m^2  \right] \\
 \l M_{\rm h, F1}^{(2)} \r^{-1}  &= - \frac{J^2}{t} m^2 \pi^2 \frac{3}{8}  \l 12 \pi + \pi^2 - 24 \r \\
 \l M_{\rm h, F2}^{(2)} \r^{-1}  &= - \frac{J^2}{t} \frac{\pi}{4096} (400 \pi + 19 \pi^3 - 1600) \\
 \l M_{\rm h, U}^{(2)} \r^{-1} &= - \frac{J^2}{t} |C|^4 \times 43.16.
\end{flalign}
By combining these results, we arrive at Eq.~\eqref{eq:MhPiComp}.

\subsubsection{Divergent integrals in the $t-J$ model}
In our discussion of the holon-polaron so far, we restricted ourselves to the $t-J^*$ model where next-nearest neighbor holon hopping is included. We can repeat our perturbative analysis for the simpler $t-J$ model, where the last three terms in Eq.~\eqref{eq:HamHolonPolaronLLP}, corresponding to next-nearest neighbor hopping, are discarded. In this case we obtain
\begin{equation}
E_{\rm h}(p_{\rm h}) = -2t \cos(p_{\rm h}) + J \l \log(2) - \nicefrac{1}{4} - \pi^2 m^2 \r,
\end{equation}
plus terms of orders $\mathcal{O}(J^2/t)$.

The second order expressions $\mathcal{O}(J^2/t)$ involve divergent integrals when the $t-J$ model is used. For example, the  Fr\"ohlich type terms from Eq.~\eqref{eq:Hfrohlich} give rise to an energy correction
\begin{equation}
E^{(2)}_{\rm h, F1}(p_{\rm h}=0) = - \frac{J^2}{t} m^2 \pi^2 \int_0^{\pi/2} dq ~ \frac{q}{1-\cos(q)}
\end{equation}
for the $t-J$ model. This expression diverges logarithmically with the infrared cut-off $\Lambda_{\rm IR}$,
\begin{equation}
E^{(2)}_{\rm h, F1}(p_{\rm h}=0) \simeq - \int_{\Lambda_{\rm IR}} dq ~ \frac{1}{q} \simeq \log \Lambda_{\rm IR}.
\end{equation}
In a finite-size system, $\Lambda_{\rm IR} = 1/L$ with $L$ the system size, and $E^{(2)}_{\rm h, F1}(p_{\rm h}=0) \simeq - \log L$ is weakly divergent.

Similarly we find that the correction to the effective mass from two-particle excitations is logarithmically divergent in the $t-J$ model, $1/M_{\rm h,F2}^{(2)} \simeq - \log L$. On the other hand, the energy $E_{\rm h, F2}^{(2)}$ is convergent. The Umklapp terms lead to a power-law divergence of the holon-polaron ground state energy,
\begin{equation}
E_{\rm h,U}^{(2)}(p_{\rm h}=0) \simeq - \frac{J^2}{t} \int dQ \frac{p(Q)}{1 + \cos Q} \simeq -\int dQ \frac{p(\pi)}{(Q-\pi)^2}
\end{equation}
in the $t-J$ model.


\begin{thebibliography}{91}
\bibitem [{fss()}]{fss}%
  \BibitemOpen
  \href@noop {} {}\bibinfo {note} {To perform the finite size scaling, we
  calculated the energy of a single hole for system sizes $L=10, 12,..., 20$
  and used a fit up to quadratic order $1/L^2$.}\BibitemShut {Stop}%
\end{thebibliography}

%

\end{document}